\documentclass[twocolumn,pra,superscriptaddress]{revtex4-1}
\usepackage{epsfig}
\usepackage{graphicx}

\usepackage[caption=false]{subfig}
\usepackage{amsmath}
\usepackage{amssymb}
\usepackage{epstopdf}
\usepackage{color}
\usepackage{epstopdf}
\usepackage{bm}
\usepackage{mathtools}
\usepackage{hyperref}
\usepackage{dsfont}
\usepackage{braket}
\usepackage{todonotes}

\begin{document}

\title{Creating cat states in one-dimensional quantum walks using delocalized initial states}
\author{Wei-Wei Zhang}
\email{weiwei.zhang@ucalgary.ca}
\affiliation{State Key Laboratory of Networking and Switching Technology, Beijing University of Posts and Telecommunications, Beijing 100876, China}
\affiliation{Institute for Quantum Science and Technology, and Department of Physics and Astronomy, University of Calgary, Canada, T2N~1N4}

\author{Sandeep K.~Goyal}
\email{skgoyal@iisermohali.ac.in}
\affiliation{Institute for Quantum Science and Technology, and Department of Physics and Astronomy, University of Calgary, Canada, T2N~1N4}

\author{Fei Gao}
\email{gaofei\_bupt@hotmail.com}
\affiliation{State Key Laboratory of Networking and Switching Technology, Beijing University of Posts and Telecommunications, Beijing 100876, China}

\author{Barry C.~Sanders}
\email{sandersb@ucalgary.ca}
\affiliation{Institute for Quantum Science and Technology, and Department of Physics and Astronomy, University of Calgary, Canada, T2N~1N4}
\affiliation{Hefei National Laboratory for Physical Sciences at Microscale, University of Science and Technology of China, Hefei, Anhui 230026, China}
\affiliation{Shanghai Branch, CAS Center for Excellence and Synergetic Innovation Center in Quantum Information and Quantum Physics, University of Science and Technology of China, Shanghai 201315, China}
\affiliation{Program in Quantum Information Science, Canadian Institute for Advanced Research, Toronto, Ontario M5G 1Z8, Canada}

\author{Christoph Simon}
\email{csimo@ucalgary.ca}
\affiliation{Institute for Quantum Science and Technology, and Department of Physics and Astronomy, University of Calgary, Canada, T2N~1N4}

\begin{abstract}
Cat states are coherent quantum superpositions of macroscopically distinct states and are useful for understanding the boundary between the classical and the quantum world. Due to their macroscopic nature, cat states are difficult to prepare in physical systems. We propose a method to create cat states in one-dimensional quantum walks using delocalized initial states of the walker.
 Since the quantum walks can be performed on  any quantum system, our proposal enables a platform-independent realization of the cat states.
We further show that the  linear dispersion relation of the effective quantum walk Hamiltonian, which governs the dynamics of the delocalized states, is responsible for the formation of the cat states. We analyze the robustness of these states against environmental interactions and present methods to control and manipulate the cat states  in the photonic implementation of quantum walks.
\end{abstract}
\maketitle

\section{Introduction}
Schr\"odinger cat states can be defined as quantum superpositions of macroscopically distinct states of a quantum system~\cite{Schrodinger1935,Milburn1986a,Yurke1986,Buzek1995}. Due to their macroscopic nature, cat states play an important role in fundamental tests of quantum theory and precision measurements~\cite{Sanders1992,Munro2002,Wenger2003,Stobinska2007}. Numerous attempts are being made to  prepare cat states in various physical systems~\cite{Monroe1996,Brune1996,Slosser1995,Jeong2004,Huang2006,Rao2011,Csire2012,Lee2012,Wu2013,Lau2014,Fischer2015,Wang2015,Friedman2000,Leibfried2005,Ourjoumtsev2007,Lvovsky2013,Bruno2013,Vlastakis2013,Wang2016}.

The  macroscopic superposition, which makes the cat states interesting also makes them hard to create in physical systems. This is because of the difficulty in controlling the evolution of macroscopic quantum systems while preserving the coherence in the state. Quantum walks inherently involve the coherent evolution of a macroscopic system.

In a quantum walk process, a quantum walker propagates on a lattice where the propagation is conditioned over its internal states (the coin states)~\cite{Venegas-Andraca2012,Kempe2003}. The quantum walker, unlike its classical counterpart, preserves the coherence during the propagation which results in a faster spread of the walker over the lattice as compared to  the classical random walks.
Quantum walks have been extensively studied to devise quantum algorithms~\cite{Ambainis2003,Shenvi2003,Childs2004,Childs2009a,Lovett2010} and to simulate various quantum phenomena~\cite{Klafter1980,Barvik1987,Strauch2006,Strauch2007,Bracken2007,Engel2007,Lee2007,Mohseni2008,Kurzynski2008,Childs2009,Kitagawa2010,Schreiber2010,Schreiber2012,Chandrashekar2010,Berry2012,Kitagawa2012,Kitagawa2012a,Asboth2012,Moulieras2013,Obuse2015,Edge2015}.

Here we propose a method to prepare the cat states in a one-dimensional discrete time quantum walk (DTQW) using delocalized initial states of the walker.
The quantum walks can be implemented on virtually any quantum system that meets the requirements (a lattice and a coin). Thus, our proposal provides a platform-independent methods to create cat states, which enables us to test the fundamental theories on more accessible  systems.

In Ref.~\cite{Cardano2015}, Cardano  et~al.~implemented a one-dimensional quantum walk on the orbital angular momentum (OAM) space of a single photon, following the proposal of Refs.~\cite{Zhang2010,Goyal2013}. In this experiment, they demonstrated that the state of the walker, which is delocalized initially, evolves to form a bimodal distribution that resembles a cat state.
Their experimental finding, which is consistent with their numerical calculations, motivates the research to find the cause of the formation of cat states and analysis of the stability of these states against the decoherence in quantum walks.

Here we start with a Gaussian (delocalized) initial state
and prove that it  evolves to form  a cat state. We clarify the conditions for the formation of the cat states for the entire range of the parameter $\theta$, which characterizes the bias in the coin flip in the quantum walk. The linearity of the dispersion relation of the low-momentum effective Hamiltonian, which governs the dynamics of the delocalized states, is shown to be the reason for the formation of the cat states in the one-dimensional quantum walks.
 Furthermore, experimentally viable methods are proposed to demonstrate the coherence in the presence of environmental interactions. Our analysis of the effects of decoherence on the quality of the cat states show that large separations in the cat states are possible even in the presence of noise.
Finally, we provide a method to stabilize and manipulate the cat states over the OAM of light.

The structure of the article is as follows: we provide the relevant background regarding the one-dimensional quantum walks in Sec.~\ref{Sec:Background}. In Sec.~\ref{Sec:CatState} and~\ref{Sec:Ana} we present our numerical and analytical findings. We discuss the effect of decoherence on the cat states in Sec.~\ref{Sec:Decoherence}. Method to control and manipulate the cat states are presented in Sec.~\ref{Sec:Control}.  We conclude in Sec.~\ref{Sec:Conclude}.

\section{Background}
\label{Sec:Background}

In this section, we present the relevant background of the one-dimensional quantum walks. We describe the regular coined quantum walks on a one-dimensional lattice, its generalization and the Hamiltonian, which governs the dynamics of the quantum walks. We conclude the section with an optical implementation scheme where the walk is performed over the OAM of a light beam.

\subsection{One-dimensional discrete time quantum walks}
\label{Subsec:DTQW}
 In a one-dimensional DTQW the walker propagates on a one-dimensional lattice.  The movements of the walker on the lattice are conditioned over the state of a two-state quantum coin. Each step in the walk consists of a coin-flip ($C$) followed by the conditional propagation ($S$). If $\{\Ket{\uparrow},\,\Ket{\downarrow}\}$ represents a set of two orthogonal states of the coin then the coin-flip operator $C$ reads~\cite{Goyal2015}
\begin{equation}
C= \big(\cos\theta \Ket{\uparrow}  +\sin\theta \Ket{ \downarrow}  \big)\Bra{ \uparrow } + \big(\sin\theta \Ket{\uparrow} - \cos\theta\Ket{\downarrow} \big)\Bra{ \downarrow},\label{Eq:Coin}
\end{equation}
where the parameter $\theta \in [0 , 2\pi)$.
The conditional propagator $S$ instructs the walker to move forward ($F = \sum_x\ket{x+1}\bra{x}$) or backward ($F^\dagger$) on the lattice conditioned over the states of the coin,
\begin{align}
S= F\otimes\left| \uparrow \right \rangle \left \langle \uparrow\right |+F^{\dagger}\otimes\left|\downarrow\right\rangle\left\langle\downarrow\right|.\label{Eq:Shift}
\end{align}
Here $x$ is the index for the lattice sites. Thus, the quantum walk propagator $Z$ reads
\begin{equation}
Z =  S(\mathds{1}\otimes C). \label{Eq:Z}
\end{equation}
Repeated action of the propagator $Z$ gives rise to the quantum walk dynamics.

One-dimensional DTQW has been generalized to simulate various dynamics. One of the most interesting generalizations is where a phase, which is linear in the position, is introduced after every step of the quantum walk~\cite{Genske2013,Cedzich2013}. The operator $F_{\rm m}$ which gives the site-dependent phase reads
\begin{equation}
  \label{Eq:MomentumShift}
  F_{\rm m} = \sum_x\exp(i\Phi x) \ket{x}\bra{x},
\end{equation}
where $\Phi$ is an independent parameter. The subscript m in the operator $F_{\rm m}$ is just a reminder that the operator $F_{\rm m}$ is a shift operator in the momentum space.
The propagator for the generalized quantum walk reads
\begin{equation}
  \bar Z  = F_{\rm m} Z.
\end{equation}

This generalized quantum walk demonstrates various interesting properties such as Bloch oscillations and quasi-periodic dynamics~\cite{Cedzich2013}. If the strength of the parameter $\Phi$ is set to be $\Phi = 2\pi/p$, where $p$ is a positive integer, then the walker recovers its original state after $2p$ number of steps for odd $p$ and after $p$ number of steps for even $p$. This feature can be used to restrict the spread of the walker on the lattice.

\subsection{Quantum walk Hamiltonian}
The Hamiltonian $H$ that governs the quantum walk dynamics can be calculated by substituting
\begin{equation}
  \label{Eq:QWalkHamil}
  Z = \exp(-iH\delta t),
\end{equation}
where $\delta t$ is the duration of a single step in the quantum walk. Here we have taken $\hbar \equiv 1$.

From the definition of the conditional propagator $S$ in~\eqref{Eq:Shift} and the operator $F$,  we can assert that the propagator $Z$ and the Hamiltonian $H$ are translation invariant. Thus, the Hamiltonian $H$ can be block diagonalized in the  momentum (or Fourier transform) basis $\{\ket{k}\}$
\begin{equation}
  H = \bigoplus_{k \in [-\pi,\pi)} H(k).
\end{equation}
Here we have considered a lattice of size $N$ with periodic boundary condition, where $N$ is taken to be much larger than the number of quantum walk steps. The variable $k$ represents the (quasi-) momentum that can take discrete values between $-\pi$ and $\pi$ in the integer multiples of $2\pi/N$.

The Hamiltonian $H(k)$ in the momentum basis can be calculated by expanding  the position eigenstates $\{\ket{x}\}$ in the momentum basis $\{\ket{k}\}$ as
\begin{equation}
  \Ket{x}  = \frac{1}{\sqrt{N}}\sum_k \exp(ikx) \ket{k}.\label{Eq:FTxk}
\end{equation}
By substituting Eq.~\eqref{Eq:FTxk} into the definition of the propagator $Z$ and using Eq.~\eqref{Eq:QWalkHamil} we arrive at
\begin{equation}
  \label{Eq:QWalkHamil2.1}
  H(k) = {\bm h}(k)\cdot {\bm \sigma}.
\end{equation}
Here ${\bm \sigma}$ is the vector $(\sigma_x,\sigma_y,\sigma_z)$ of Pauli spin matrices and ${\bm h}(k) = (h_1(k),h_2(k),h_3(k))$ is a three dimensional real vector, which reads
\begin{align}
h_{1}(k)&=-R(k)\sin\theta  \cos k,\\
h_{2}(k)&=R(k)\sin\theta \sin k,\\
h_{3}(k)&=-R(k)\cos\theta \cos k,\\
R(k) &= \frac{\cos^{-1}(-\cos \theta \sin k)}{\sqrt{\sin^2\theta  \sin^2k+\cos^2k}}.
\end{align}

Interestingly, for small values of the parameter $\theta$ and small $k$, the Hamiltonian $H(k)$ takes a special form that resembles a two-component Dirac Hamiltonian  (see Appendix~\ref{App:Dirac}). In this limit the effective Hamiltonian, which we represent by $H_{\rm d}$, reads
\begin{align}
H_{\rm d}(k)&= -\left( k + \frac{\pi}{2}\right)\sigma_z -\theta\frac{\pi}{2} \sigma_x.\label{Eq:DiracHamil}
\end{align}
In the Hamiltonian $H_{\rm d}(k)$ the parameter $\theta$ characterizes the mass of the particle.

\subsection{Implementing quantum walks in optical system}
\label{Sub:Implementation}
In this section, we describe an implementation scheme to realize a one-dimensional quantum walk on the OAM of light. This scheme was proposed in~\cite{Goyal2013} and experimentally demonstrated in~\cite{Cardano2015}. The purpose of this section is to familiarize the readers with an implementation scheme for the cat states in the quantum walks.
Using this implementation for the one-dimensional quantum walks we will propose a method to manipulate and control the cat states.

In this implementation scheme,  the OAM of light serves as the lattice and the polarization is used as the coin. The conditional propagator $S$~\eqref{Eq:Shift} is constructed by means of a q-plate which is a device that couples the OAM of light with its  spin angular momentum (polarization)~\cite{Marrucci2006}. The action of a q-plate on the combined state of the OAM and the polarization is given by
\begin{align}
  \label{Eq:q-plate}
  \ket{L,\ell} &\to \ket{R,\ell-2q}, \\
 \ket{R,\ell}  &\to  \ket{L,\ell+2q},
\end{align}
where $\ket{L}$ and $\ket{R}$ are the left- and right-handed circular polarization of light, and $\ket{\ell}$ is the OAM state  that has angular momentum proportional to   $\ell \hbar$. The half-integer parameter $q$ characterizes the q-plate.

A half-wave plate with its fast axis parallel to the horizontal axis interchange the left- and right-handed circular polarization. Therefore, a q-plate with $q=1/2$ followed by a half-wave plate give rise to the conditional propagator $S$~\eqref{Eq:Shift}.

The coin-flip operator $C$~\eqref{Eq:Coin} can be implemented using the Simon-Mukunda polarization gadget~\cite{Simon1989}. This gadget is a combination of one half-wave plate and two quarter-wave plates, and can be used to realize an arbitrary SU$(2)$ operation on the polarization of light. Hence, the  quantum walk propagator $Z$ can be simulated using a q-plate, a half-wave plate, and a Simon-Mukunda polarization gadget in series. Placing these three components in a loop can realize a one-dimensional quantum walk on the OAM of light.

\section{Cat states in quantum walks}
\label{Sec:CatState}

\begin{figure}
  \includegraphics[width=0.45\textwidth]{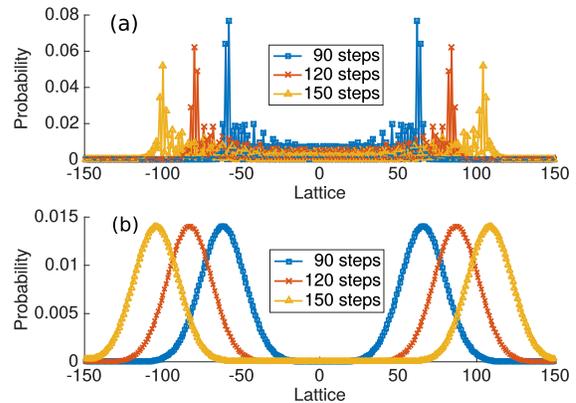}
  \caption{The evolution of the walker on a one-dimensional lattice for (a) localized and (b) delocalized initial states. Here, the parameter $\theta = \pi/4$, and the width of the Gaussian for the  figure (b) is $\sigma \approx 10$.
  }
  \label{Fig:QWalk}
\end{figure}

In this section, we demonstrate the formation of the cat states in the one-dimensional DTQW. We show that the walker in a delocalized (Gaussian) initial state evolves to form a cat state. We present methods to analyze the cat nature of the evolved state of the walker.

Quantum walk evolution, typically, results in a bimodal distribution of the walker on the lattice. In Fig.~\ref{Fig:QWalk}, we plot the probability distribution of the walker at time $t = 90,~120,~150$ steps for localized and delocalized  initial states. In Fig.~\ref{Fig:QWalk}a, the  initial state of the walker is localized at the origin. The state of the walker evolves to a bimodal distribution with a residual probability between the two components of the distribution. The residual probability signifies the overlap between the two components of the distribution. Hence, the evolved state is not a cat state.

In Fig.~\ref{Fig:QWalk}b, we start with a delocalized initial state $\Ket{\Psi(0)}_{\rm de}$ of the walker
\begin{equation}
  \label{Eq:Delocalized}
  \Ket{\Psi(0)}_{\rm de} = \frac{1}{\mathcal{N}}\sum_n  \exp\left(-\frac{n^2}{4\sigma^2}\right) \ket{n}\otimes \Ket{\chi}_c,
\end{equation}
 which has a Gaussian probability distribution. We find that the delocalized state $\Ket{\Psi(0)}_{\rm de}$ evolves to a state $\Ket{\Psi(t)}_{\rm de}$ after time $t$ that has the bimodal probability distribution with vanishing residual probability between the two  components of the bimodal distribution.
Here we have chosen  the width  $\sigma$ of the Gaussian to be sufficiently large (about $10$ lattice sites). $\Ket{\chi}_c$ is a normalized initial state of the coin, and $\mathcal{N}$ is the normalization constant. The two components of the bimodal distribution can represent macroscopically distinct states of the walker. Hence, the evolved state can be seen as a cat state. In the remainder of this section, we analyze the conditions required for the evolved state to be a cat state.

\subsection{ Small $\theta$ case}
We start with a simple case when the parameter $\theta$ is small. In this limit, the quantum walk Hamiltonian can be approximated to a two-component Dirac Hamiltonian $H_{\rm d}$~\eqref{Eq:DiracHamil}. In this limit, the quantum walk can be used to simulate quantum relativistic effects such as Klein paradox and Zitterbewegung~\cite{Strauch2007,Kurzynski2008}. Thus, this limit can be considered as the relativistic limit of the quantum walk.

 The parameter $\theta$ in the Dirac Hamiltonian $H_{\rm d}$ characterizes the mass of the particle. For $\theta = 0$ the Hamiltonian $H_{\rm d}$ represents a massless particle. If the initial state of the walker in the momentum space is
 \begin{equation}
   \ket{\Psi(0)} = \sum_k\ket{\psi_k}\otimes \left(a \ket{\uparrow} + b \ket{\downarrow}\right),
 \end{equation}
then the evolved state, for the case $\theta = 0$, reads
\begin{align}
  \ket{\Psi(t)} &= \exp(-iH_{\rm d} t) \ket{\Psi(0)},\nonumber\\
&= \sum_k\left(ia e^{ikt}\ket{\psi_k}\otimes \ket{\uparrow} - ib e^{-ikt}\ket{\psi_k}\otimes \ket{\downarrow}\right).\label{Eq:DiracEvo}
\end{align}
Here, the two orthogonal spin components of the particle propagate  in the opposite directions independent of each other. Due to the linear dispersion relation in the Dirac Hamiltonian, the evolution does not result in the spreading of the wave function of the particle, which results in the formation of cat states.

The same feature, namely, the non-dispersive behaviour of the wave function, persists for non-zero values of $\theta$ as long as $\theta$ is small. Thus, cat states can be formed in the relativistic limit of the one-dimensional quantum walks.

\subsection{ Arbitrary $\theta$ case}
\begin{figure}
  \includegraphics[width=0.45\textwidth]{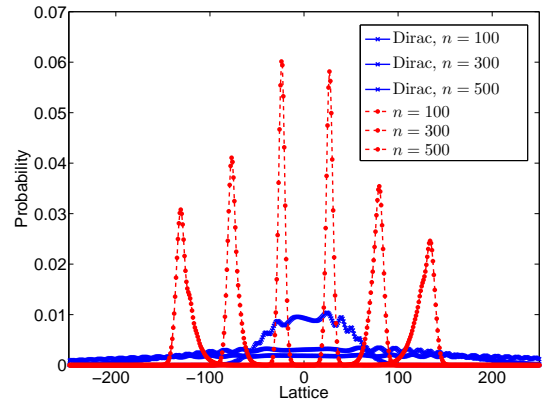}
\caption{The comparison between the exact quantum walk evolution (the dashed line) and the evolution using the Dirac Hamiltonian (the solid line) for large values of $\theta$ ($\theta = \pi/2.4 = 75^\circ$). This figure shows the spread in the width of the Gaussian in the case of the Dirac evolution but almost no spread in the exact quantum walk evolution.
} \label{Fig:Dirac}
\end{figure}

In the limit when $\theta$ is large, the Dirac description of the quantum walk breaks down, therefore, one might not expect to observe the cat states.
In Fig.~\ref{Fig:Dirac}, we plot the probability distribution of the walker over the lattice at different times. Here we have considered two different dynamics for the walker, one where we use the exact quantum walk evolution to propagate the walker on the lattice and other where we use Dirac Hamiltonian to propagate the walker. We have chosen $\theta = \pi/2.4$, i.e., a large value of $\theta$. From this figure, we see that the Dirac Hamiltonian and the exact quantum walk dynamics result in strikingly different evolutions. The cat-state-like distribution persists for large $\theta$ in the exact quantum walk evolution where Dirac description predicts only dispersed wave function.

In the following, we show that the evolved states achieved for an arbitrary $\theta$ and the state $\Ket{\Psi(t)}$~\eqref{Eq:DiracEvo} achieved in the small $\theta$ limit are qualitatively the same.
In order to see that, first, we notice that the state $\Ket{\Psi(t)}$ in Eq.~\eqref{Eq:DiracEvo} is highly entangled and the wave-packets corresponding to the orthogonal states of the coin are non-dispersive and propagate in the opposite directions.

\begin{figure*}
\subfloat[  \label{Fig:Entanglement}]
 { \includegraphics[width=0.33\textwidth]{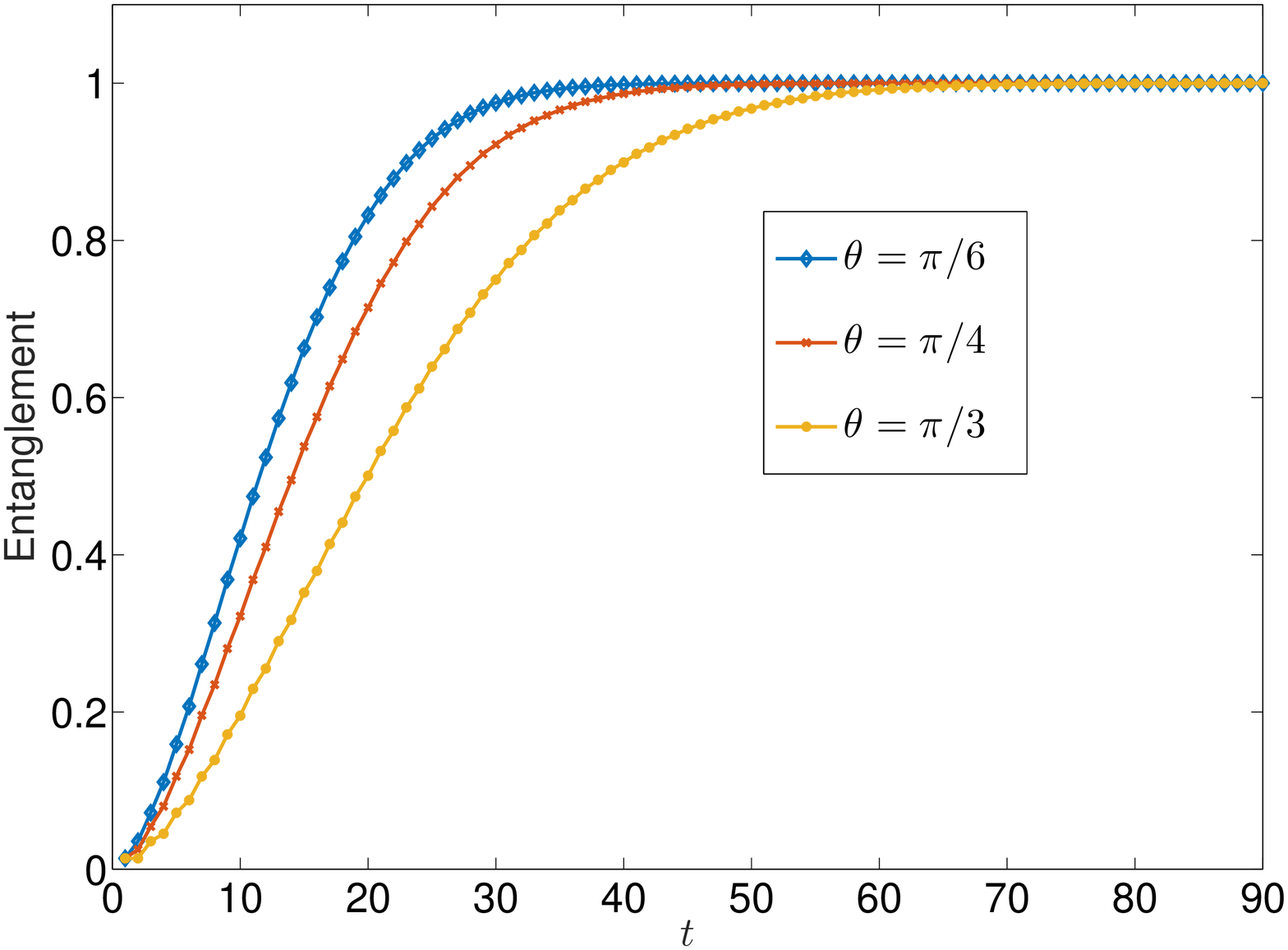}}~
\subfloat[  \label{Fig:SchmidtGaussian}]
{ \includegraphics[width=0.33\textwidth]{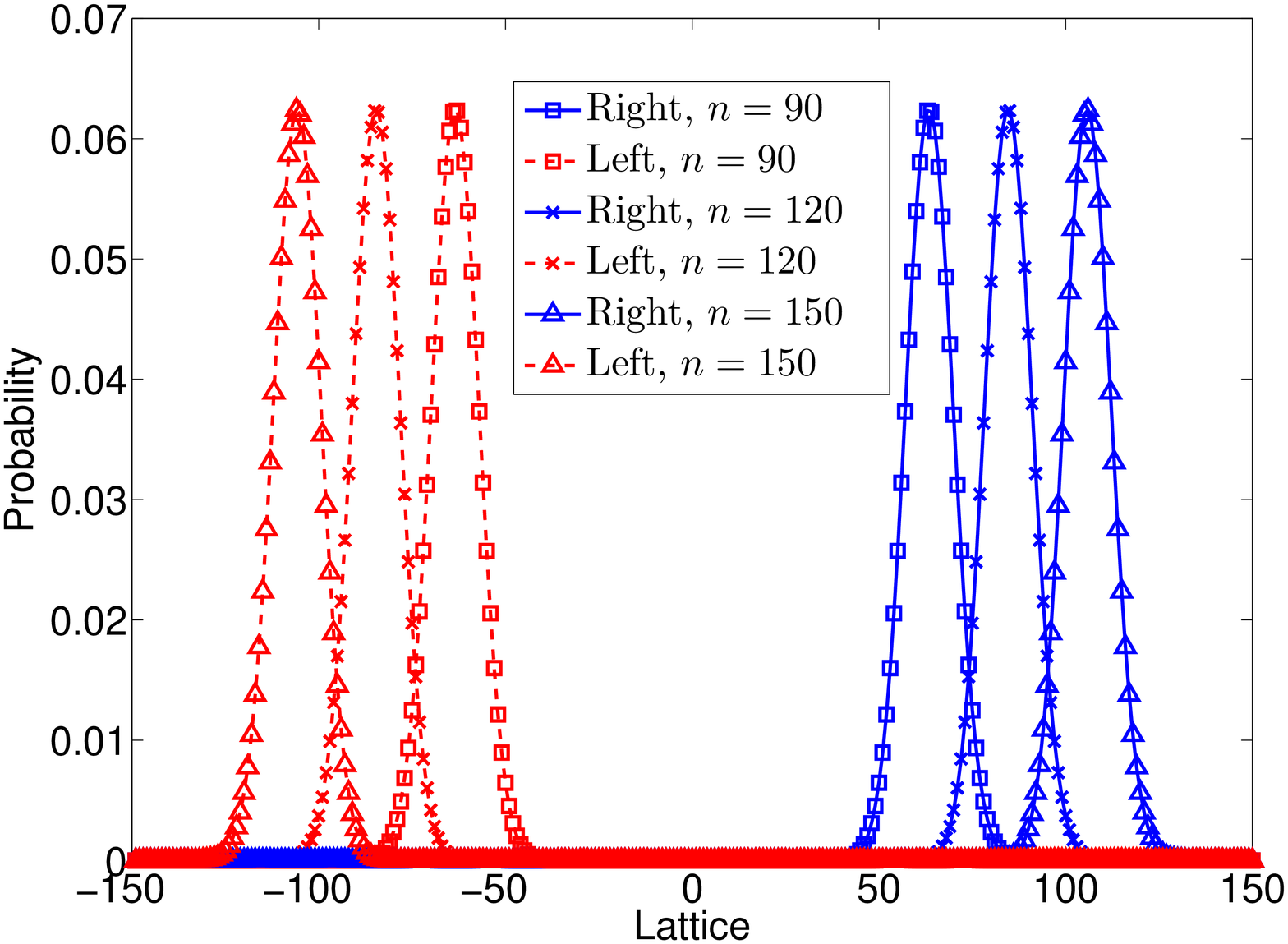}}~
\subfloat[\label{Fig:WidthGaussian}]
{\includegraphics[width=0.33\textwidth]{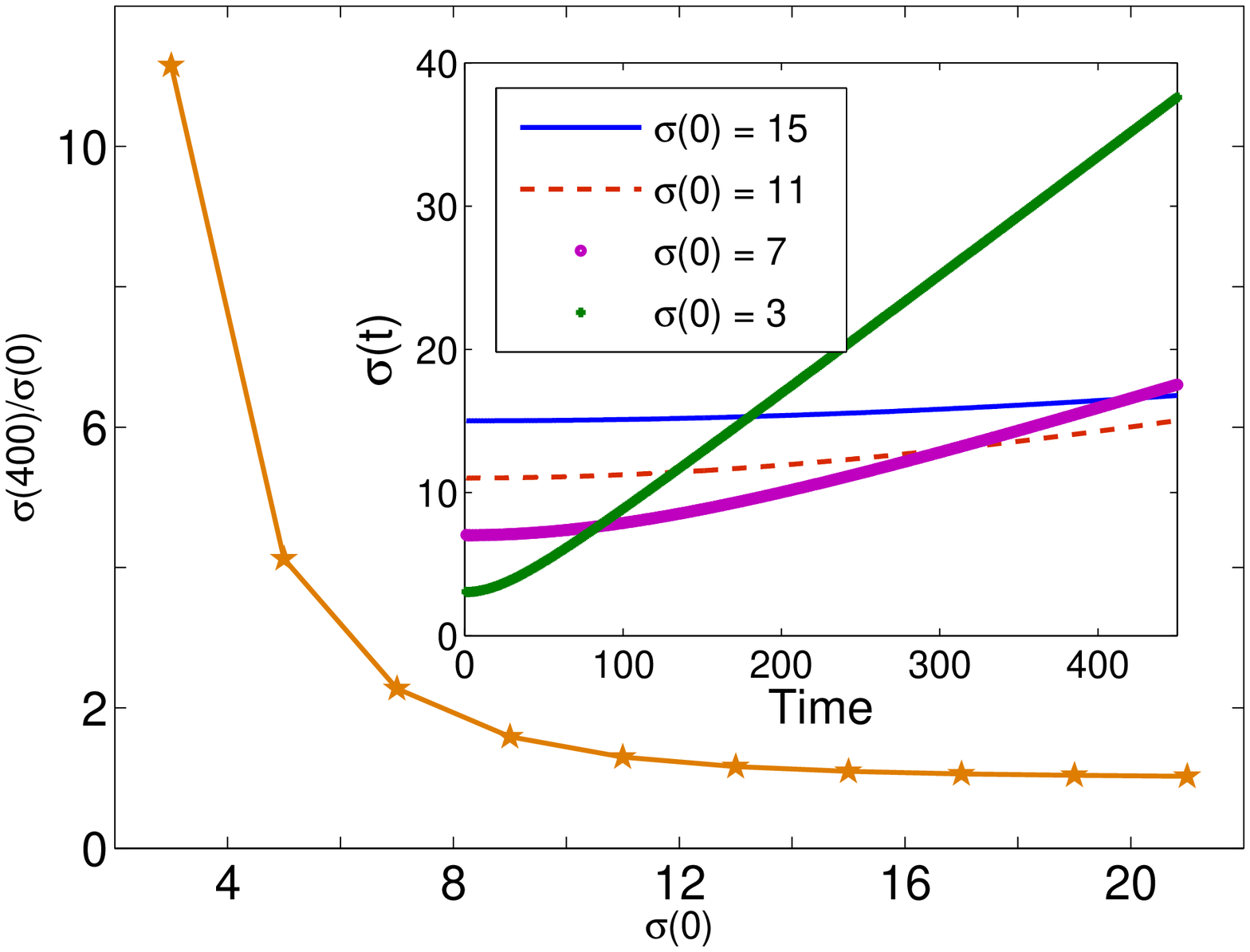}}
\caption{In this figure we summarize the numerical evidence in favour of the cat states in one dimensional DTQW. Here we have chosen the parameter $\theta = \pi/4$ and the width of the Gaussian $\sigma \approx 10$, unless specified explicitly.
 In Fig.~(a) we plot the entanglement between the coin and the lattice as a function of time for different values of $\theta$. The entanglement is calculated by calculating the von-Neumann entropy of the reduced density matrix of the coin. In Fig.~(b) we show the macroscopically distinct states of the walker propagating in the opposite directions. The two macroscopically distinct states correspond to the states $\Ket{X}$ and $\Ket{X_\perp}$ introduced in Eq.~\eqref{Eq:EvolvedState}. 
 Here the solid curves are the Gaussian moving towards the right and the dashed curves are the Gaussian moving towards the left. (c) Here we plot the normalized width of the Gaussian after $400$ steps, i.e., $\sigma(400)/\sigma(0)$ against the initial width of the Gaussian $\sigma(0)$. In the inset we show the time evolution of the width of the Gaussian for different initial width. Clearly, as we increase the width of the initial wave-packet the normalized width of the Gaussian at large times saturates to the value $1$. In other words, increasing the initial width suppresses the spreading of the wave-packet.}\label{Fig:CatStates}
\end{figure*}

In Fig.~\ref{Fig:Entanglement} we plot the entanglement in the state $\Ket{\Psi(t)}_{\rm de}$ between the coin and the walker. The entanglement is calculated by first calculating the reduced density matrix of the coin (or the walker) and then calculating the von-Neumann entropy of the reduce density matrix~\cite{Nielsen2010}. In this figure, we can see that the entanglement approaches the maximum value after sufficiently long time.
In Fig.~\ref{Fig:SchmidtGaussian} we plot the  distribution for the states of the walker corresponding to the two orthogonal states of the coin, which are calculated by diagonalizing the reduced density matrix of the coin. Clearly the two wave-packets are moving in the opposite directions.
 The maximum entanglement along with the purity  show that the state $\Ket{\Psi(t)}_{\rm de}$ must have the form
\begin{equation}
  \label{Eq:EvolvedState}
  \Ket{\Psi (t)}_{\rm de} = \frac{1}{\sqrt{2}} \left(\Ket{X(t)}\otimes \ket{\phi(t)} + \Ket{X_\perp(t)} \otimes \ket{\phi_\perp(t)}\right),
\end{equation}
where the states $\Ket{X(t)}$ and $\ket{X_\perp(t)}$ represent the two non-overlapping wave-packets and $\ket{\phi(t)}$ and $\ket{\phi_\perp(t)}$ are the orthogonal states of the coin.

In Fig.~\ref{Fig:WidthGaussian}, we demonstrate the effect of the width of the initial state on the spreading of the wave-packet. Here we plot the normalized width of the evolved Gaussian wave-packet $\sigma(t)/\sigma(0)$ at a very large time $t = 400$ steps as a function of the initial width $\sigma(0)$. This plot shows that the normalized width of the Gaussian approaches the value $1$ as we increase the width of the initial wave-packet. In the inset we plot the time evolution of the width of the Gaussian wave-packets $\sigma(t)$. We choose $\theta = \pi/4$ and the width of the initial Gaussian to be $\sigma = 3,~7,~11,~15$. These two plots combined confirm  that the spreading of the wave-packet decreases as the width of the initial state is increased.

\begin{figure}
{\includegraphics[width=0.45\textwidth]{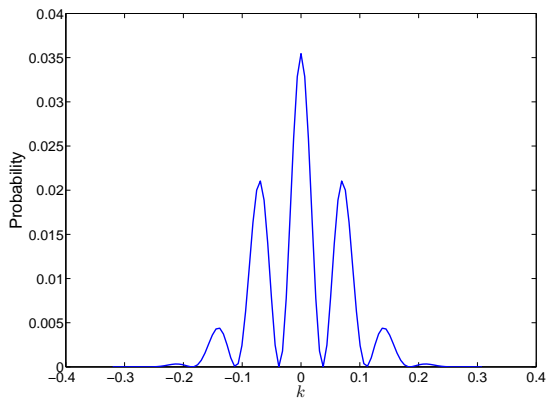}}
\caption{The probability distribution in the momentum space after projecting the evolved state on a chosen coin state. Here we have chosen $\theta = \pi/4$ and the coin states $\ket{\chi}_c = \ket{\chi'}_c = \ket{u_-(0)} + i\ket{u_+(0)}$, where $\ket{u_\pm(0)}$ are the eigenvectors of the quantum walk Hamiltonian corresponding to $k=0$. The occurrence of the fringes in this distribution signifies the coherence between the two macroscopic states $\Ket{X}$ and $\Ket{X_\perp}$ of the walker. }  
\label{Fig:CatFourier}
\end{figure}

Another method to verify the coherence in the two wave-packets in the evolved state $\Ket{\Psi(t)}_{\rm de}$ is by studying the probability distribution of the walker in the momentum space after projecting over an appropriate state of the coin. This can be done as follows:
if the states $\Ket{X(t)}$ and $\Ket{X_\perp(t)}$ are coherent Gaussian states that have the form
\begin{equation}
 \Ket{G(\pm n_t,\sigma)} =\frac{1}{\mathcal{M}}\sum_n \exp\left(-\frac{(n\pm n_t)^2}{4\sigma^2}\right) \ket{n},
\end{equation}
with the mean at $\pm n_t$ and the width $\sigma$, then the Fourier transform of these states read
\begin{equation}
  \Ket{G(\pm n_t,\sigma)} \to \frac{1}{\mathcal{M}}\sum_k e^{\mp in_tk}e^{-\sigma^2k^2/2}\ket{k}.
\end{equation}
Thus, the state $\Ket{\Psi(t)}_{\rm de}$ in the momentum basis reads
\begin{equation}
  \Ket{\tilde\Psi_t} = \frac{1}{\sqrt{2}\mathcal{M}} \sum_k e^{-\sigma^2k^2/2}\ket{k} \otimes\left(e^{- in_tk} \ket{\phi} + e^{in_tk} \ket{\phi_\perp}\right).
\end{equation}
 After projecting the state $\Ket{\tilde\Psi_t}$ on the coin state $\Ket{\chi'}_c$, the state of the walker reads
\begin{align}
  \ket{\Psi} &=  \frac{1}{\sqrt{2}\mathcal{M}} \sum_k e^{-\sigma^2k^2/2}\left(\alpha e^{- in_tk} + \beta e^{in_tk} \right)\ket{k},\label{Eq:ProjCoin}
\end{align}
where
\begin{align}
  \alpha &= \prescript{}{c}{\Braket{\chi'|\phi(t)}},\quad  \beta = \prescript{}{c}{\Braket{\chi'|\phi_\perp(t)}}.
\end{align}
Note that, $\Ket{\Psi}$ in~\eqref{Eq:ProjCoin} represents a state of the walker which is a superposition of two Gaussians in the position space centred around $\pm n_t$. Thus, the state $\Ket{\Psi}$ itself is a cat state as it contains a coherent superposition of two macroscopically distinct states.

 For $\alpha = \beta$, the probability distribution corresponding to $\Ket{\Psi}$ in the momentum space will be a product of a Gaussian and $\cos^2n_tk$. For an appropriate choice of $\ket{\chi'}_c$ one can acquire $\alpha = \beta$.  Thus,  the presence of the fringes in the momentum space probability distribution signifies the coherence in the two Gaussian probability distributions in the evolved state of the quantum walk.
In Fig.~\ref{Fig:CatFourier} we plot the probability distribution for the state $\ket{\Psi}$ in the momentum space. The clear presence of the fringes in the plot ensures that the two Gaussian probability distributions in the evolved state of the walker are coherent, thus; the evolved state is a cat state.

\begin{figure}
  \includegraphics[width=0.45\textwidth]{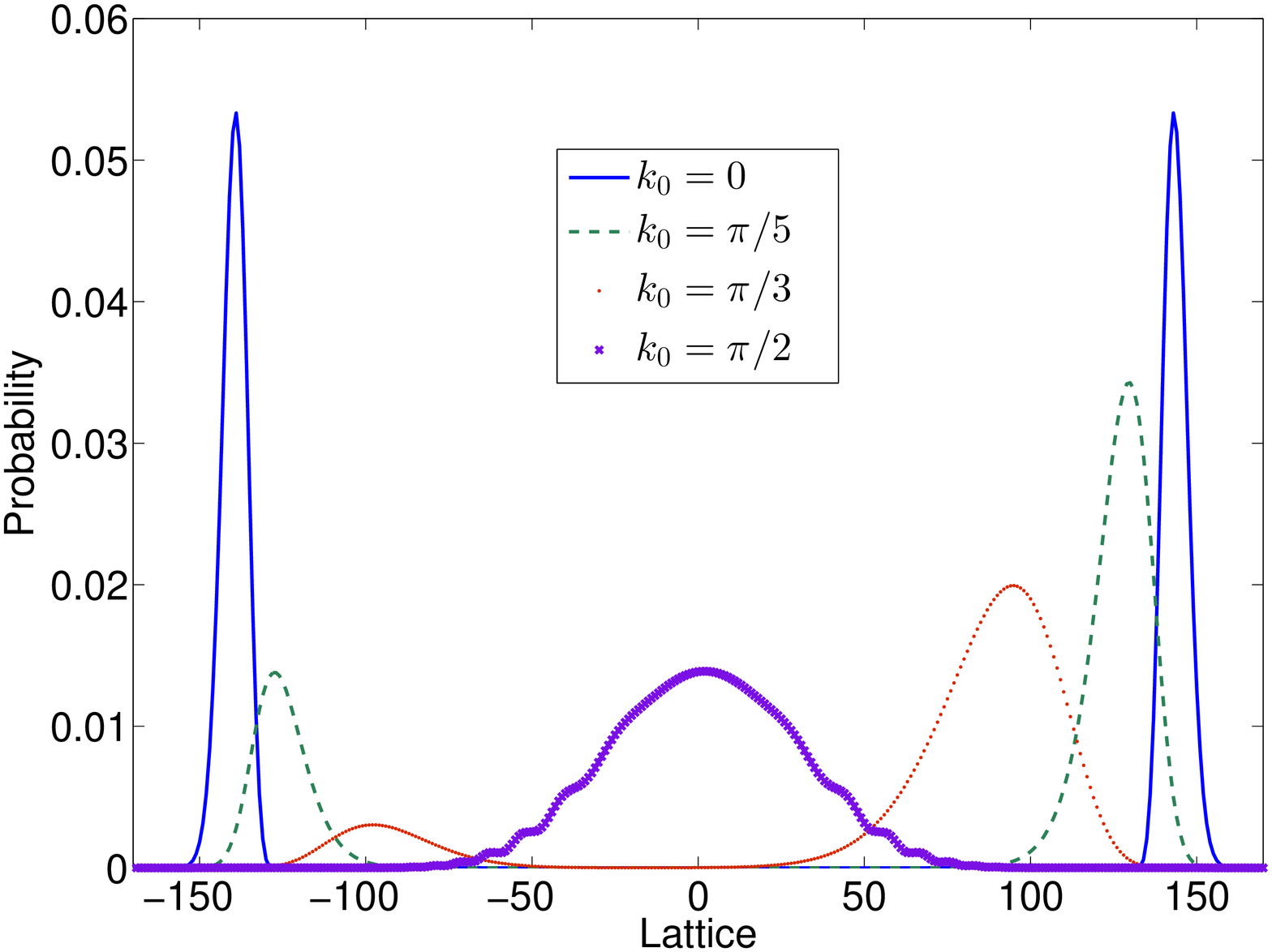}
  \caption{The plot for the probability distribution of the evolved state for different values of  the mean momentum $k_0$. Here we have set $\theta = \pi/4$ and the width of the Gaussian $\sigma = 10$.}
  \label{Fig:ReturnK0}
\end{figure}

Until now we have considered only those cases when the initial state of the walker is centred around $k=0$; therefore, the average momentum of the walker is small. What happens when the initial state is a Gaussian but not centred at $k=0$? In Fig.~\ref{Fig:ReturnK0} we plot the probability distribution for different initial states. Here we consider the initial state of the walker to have a Gaussian probability distribution and the mean value of the momentum to be $0 \le k_0 \le \pi/2$. From Fig.~\ref{Fig:ReturnK0} it can be seen that we get perfect cat states only when $k_0 \approx 0$.

In this section, we have shown that the delocalized initial states of a quantum walker evolve to form the cat states. This result is independent of the coin parameter $\theta$. However, the formation of the cat states strongly depends on the mean value of the momentum in the initial state. So far our analysis was based only on numerical results. In the following section, we present the analytic description for the formation of the cat states in quantum walks for the entire range of $\theta$ including the large $\theta$ regime where the Dirac Hamiltonian does not comply.

\section{Analytic approach to cat states in quantum walks}
\label{Sec:Ana}
The numerical results, although compelling, do not give us the real physics behind the formation of the cat states in the quantum walks. In this section, we present the reasons behind the formation of the cat states in the quantum walk.

An important result in the previous section is that the cat states are formed due to the delocalized initial states that are centred around zero momentum.  It suggests that  the low-momentum behaviour of the quantum walks is responsible for the formation of the cat states. Furthermore, from the small $\theta$ limit of the quantum walk Hamiltonian, i.e., Dirac Hamiltonian, we can see that the linear dispersion relation and the momentum-independent eigenvectors of the Hamiltonian cause the formation of the cat states.

Interestingly, for small values of the momentum $k$ the Hamiltonian $H(k)$ in Eq.~\eqref{Eq:QWalkHamil2.1} also has linear dispersion even though the Hamiltonian $H(k)$ itself is non-linear in $k$ (see Appendix~\ref{App:QWalkHamil} for detailed calculations)
\begin{equation}
E_\pm(k) = \pm   \left(k\cos\theta + \frac{\pi}{2}\right) + O(k^3).\label{Eq:LinDep}
\end{equation}
In other words, the energy $E_\pm(k)$ does not have second order terms in $k$ and for small values of $k$ (say $k <\pi/20$) the $k^3$ terms can be neglected, hence, giving rise to linear dispersion relation.

Furthermore,  the eigenvectors $\Ket{u_\pm(k)}$ of the Hamiltonian $H(k)$ depend weakly on the momentum $k$ for small values of $k$ (see Appendix~\ref{App:QWalkHamil})
\begin{equation}
  |\Braket{u_{i}(0)|u_j(k)}|^2 = \delta_{ij}+O(k^2).\label{Eq:Overlap}
\end{equation}
 Eq.~\eqref{Eq:Overlap} along with the linear dispersion relation is responsible for the formation of the cat states. This can be understood as follows: if  we start with a delocalized state $\Ket{\tilde\Psi(0)}_{\rm de}$ of the walker
\begin{equation}
  \Ket{\tilde\Psi(0)}_{\rm de} =   \frac{1}{\mathcal{N}'}\sum_k  \exp\left(-\frac{k^2}{4\delta^2}\right) \ket{k}\otimes \Ket{\chi}_c,
\end{equation}
which has a Gaussian spread in the momentum space, centred around $k=0$ and having the width $\delta < \pi/20$, and the coin state $\ket{\chi}_c$, then the evolved state at time $t$ reads
\begin{align}
  &\Ket{\tilde\Psi(t)} = \frac{1}{\mathcal{N}'}\sum_k  \exp\left(-\frac{k^2}{4\delta^2}\right) \ket{k}  \otimes\nonumber\\
& \qquad\left(e^{-iE_-(k)t}a_-(k)\Ket{u_-(k)}+ e^{-iE_+(k)t}a_+(k)\Ket{u_+(k)}\right),
\end{align}
where $ a_\pm(k) = \Braket{u_\pm(k)|\chi}_c$. Now projecting the state $\Ket{\tilde\Psi(t)}$ on the coin state $\Ket{\chi}_c$ results in state of the walker
\begin{align}
  \Ket{\Psi}_{\rm mom} &= \frac{1}{\mathcal{N}'}\sum_k  \exp\left(-\frac{k^2}{4\delta^2}\right) \left(e^{-iE_-(k)t}|a_-(k)|^2\right.\nonumber\\
& \qquad\qquad\qquad \left. + e^{-iE_+(k)t}|a_+(k)|^2\right)\ket{k}.\label{Eq:Mom}
\end{align}
The state $\Ket{\Psi}_{\rm mom}$ in~\eqref{Eq:Mom} is the same as the state $\Ket{\Psi}$ in~\eqref{Eq:ProjCoin} with $\alpha = |a_-|^2,~\beta = |a_+|^2$ and $\delta = 1/\sigma$, and in the position space $\Ket{\Psi}_{\rm mom}$ represents a state which is in a superposition of two Gaussians centred around $\pm t\cos\theta$. Hence, $\Ket{\Psi}_{\rm mom}$ represents a cat state.

Alternatively, if $|a_-(k)|^2 = |a_+(k)|^2$ and independent of $k$ then the probability distribution corresponding to the state $\Ket{\Psi}_{\rm mom}$ in the momentum space is a product of a Gaussian and $\cos^2E_-(k)t$. This means the probability distribution corresponding to the state $\Ket{\Psi}_{\rm mom}$ has fringes exactly like the one in Fig.~\ref{Fig:CatFourier}. In that case the state $\Ket{\tilde\Psi(t)}_{\rm de}$ represents a cat state.

 For appropriate choices for the state $\Ket{\chi}_c$ we can get $|a_-(k)|^2 \approx |a_+(k)|^2$ which, for small values of $k$, is $k$-independent. Using Eq.~\eqref{Eq:Overlap} we construct one such class of state which reads
\begin{equation}
  \Ket{\chi}_c = \frac{1}{\sqrt{2}}\left(\Ket{u_-(0)} + e^{i\varphi}\Ket{u_+(0)}\right),
\end{equation}
where $\varphi$ is a free parameter.
This class satisfies the relation
\begin{equation}
  |a_-(k)|^2 \approx |a_+(k)|^2 \approx \frac{1}{2}.
\end{equation}
This completes our proof that the Hamiltonian $H(k)$, and hence the one-dimensional DTQW,  gives rise to the cat states.

Let us emphasize  that the linear dispersion~\eqref{Eq:LinDep} does not mean that the Hamiltonian is linear. In fact in our case, if we truncate the Hamiltonian $H(k)$ to the first order in $k$, then we will not get the linear dispersion relation for large values of the parameter $\theta$. The $O(k^2)$ terms in the Hamiltonian $H(k)$ make the dispersion relation linear. Another interesting point to note is that the energy $E_\pm(k)$~\eqref{Eq:LinDep} is a constant for two exceptional values of the $\theta$, i.e., for $\theta = \pi/2$ and $\theta = 3\pi/2$ the energy $E_\pm(k)$ is independent of the momentum $k$. Therefore, at these values of $\theta$ the quantum walk does not evolve and hence we can not realize the cat states.

To summarize, we have shown that the formation of the cat states is due to the linear dispersion relation and the weak dependence of the eigenvectors of the quantum walk Hamiltonian on the momentum $k$. In the following section, we analyze the effect of decoherence on the cat states in quantum walks.

\section{Effect of the environmental interactions on the cat states}
\label{Sec:Decoherence}

\begin{figure}
  \includegraphics[width=0.45\textwidth]{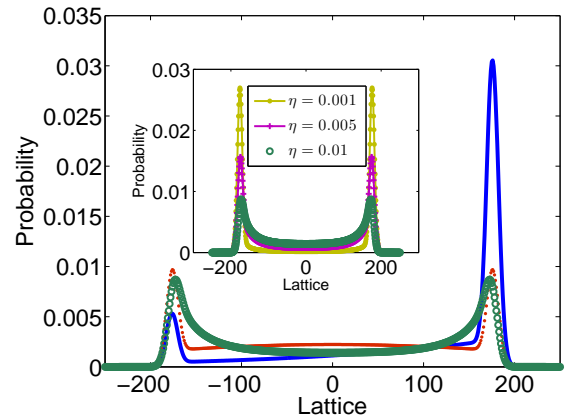}
  \caption{Spread of the walker on the lattice in the presence of amplitude damping (blue solid curve), bit-flip (red dotted curve) and pure dephasing (green circles)  after $250$ steps. Here the amplitude damping and the bit-flip baths are acting on the coin, whereas the pure dephasing type bath is acting only on the walker. The value of $\eta$ is set to be $0.01$. In the inset we plot the probability distribution of the walker after $250$ steps for different values of $\eta$.}  \label{Fig:DecoherenceProb}
\end{figure}

The discussion of the cat states is incomplete without considering the effects of the environmental interactions with the quantum system. Cat states are highly susceptible to their surroundings. Therefore,  establishing  the feasibility of forming cat states in a quantum system interacting with a bath is important. In this section, we study the effect of amplitude damping, bit-flip and pure dephasing  on the quality of the cat states. The amplitude damping and the bit-flip baths act only on the coin, whereas the pure dephasing type bath can act on the coin, on the walker and on both the coin and the walker together. We consider all the scenarios in our analysis. 

The action of a pure dephasing type bath on a given density matrix $\rho$ can be defined by the relation~\cite{Kendon2003,Kendon2007,Broome2010}
\begin{equation}
  \rho \to \tilde\rho = e^{-\eta t}\rho + (1-e^{-\eta t}){\rm diag}(\rho) = \hat{V}_{\rm d}(\rho).
\end{equation}
Here $\eta$ characterizes the strength of the bath, $\eta = 0$ implies no interaction with the bath. The function ${\rm diag}(\rho)$ keeps the diagonal elements of the matrix $\rho$ and discard all the off-diagonal elements. 

The transformation of a given density matrix $\rho$ under the influence of the amplitude damping bath and the bit-flip bath can be written using the Kraus operators $\{M_0,M_1\}$ as
\begin{equation}
  \rho \to \tilde\rho = M_0\rho M_0^\dagger +M_1\rho M_1^\dagger = \hat{V}(\rho).
\end{equation}
The Kraus operators $ \{M_0,M_1\}\equiv \{A_0,A_1\}$ for the  amplitude damping bath can be written as
\begin{equation}
  A_0 \equiv \begin{pmatrix}
    1 & 0\\ 0 & {e^{-\eta t/2}}\end{pmatrix},\quad
  A_1 \equiv \begin{pmatrix}
    0 &  \sqrt{1-e^{-\eta t}}\\0 & 0\end{pmatrix},             
\end{equation}
and $ \{M_0,M_1\}\equiv \{B_0,B_1\}$  for bit-flip bath as
\begin{equation}
B_0 \equiv {e^{-\eta t/2}} \begin{pmatrix}
    1 & 0\\ 0 & 1\end{pmatrix},\quad
  B_1 \equiv \sqrt{1-e^{-\eta t}}\begin{pmatrix}
    0 & 1 \\1 & 0\end{pmatrix}.
\end{equation}

We incorporate the effect of these baths in our evolution by applying the superoperator $\hat{V}$ (or $\hat{V}_{\rm d}$ for pure dephasing) after every step of the quantum walk. In Fig.~\ref{Fig:DecoherenceProb}, we plot the spread of the walker over the lattice in the presence of all these baths. Interestingly, we still get the bimodal distribution with an additional  residual probability between the two peaks. For the case of amplitude damping bath (the blue solid curve) the evolution is not symmetric. This is due to the fact that in amplitude damping one state of the coin is favoured over the other. Therefore, the walker prefer to move in one direction over the other. The bimodal feature of the probability distribution of the walker is preserved even for strong environmental interactions (see  inset of Fig.~\ref{Fig:DecoherenceProb}).

Although the evolved state in the presence of noise has a similar bimodal distribution as in the case of pure states (without noise), the coherence in the two cases can be very different. To quantify the coherence in the evolved state of the walker we can calculate the revival fidelity of the evolved state upon reversing the dynamics using a physical operation~\cite{Dalvit2000,Lau2014}. If the state of the walker remains pure in the evolution then the walker can regain its original state by reversing the dynamics. However, if the walker loses the purity in the evolution then the revival is not perfect.

To quantify the coherence, first, we need to devise an operation that can reverse the dynamics of the quantum walk.
In our numerical calculations, we find that the Pauli spin operator $\sigma_y$ acting on the coin state of the walker can be used to reverse the direction of propagation of the walker if the initial state of the walker is delocalized.

Using the $\sigma_y$ operator we can calculate the revival fidelity as follows: we first evolve the initial delocalized state of the walker for time $T$ in the presence of the bath. At this point, we reverse the dynamics by applying the $\sigma_y$ operator on the coin. We again evolve the state for time $T$ in the presence of the bath followed by $\sigma_y$ operation. Now we can calculate the fidelity between the evolved state $\rho(2T)$ and the initial state $\Ket{\Psi(0)}_{\rm de}$ as
\begin{equation}
  r = \prescript{}{\rm de}{\Braket{\Psi(0)|(\mathds{1}\otimes \sigma_y)\rho(2T)(\mathds{1}\otimes \sigma_y)|\Psi(0)}}_{\rm de}.
\end{equation}
High values of the revival fidelity $r$ signifies high amount of coherence in the state.

\begin{figure}
  \includegraphics[width=0.45\textwidth]{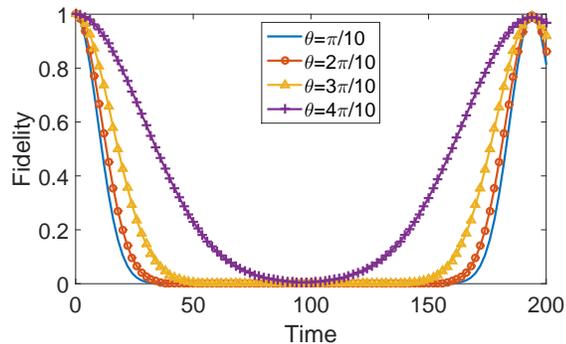}
  \caption{Plot for the revival fidelity between the evolved state and the initial state for different values of $\theta$ in the absence of the dephasing. For first $T = 97$ steps the walk is uninterrupted at which point we apply the $\sigma_y$ operation. The application of $\sigma_y$ causes the walker to retrace its footsteps resulting in a rise in the fidelity reaching the maximum in $194$ steps. Here we have plotted the values of the fidelity only for the even number of steps as the fidelity for the odd number of steps is zero.
  }
  \label{Fig:RevivalFid}
\end{figure}

In Fig.~\ref{Fig:RevivalFid} we plot the revival fidelity in the quantum walk evolution for various values of $\theta$ in the absence of noise. Here we apply the $\sigma_y$ operation after $T=97$  steps. Till then the fidelity between the evolved state and the initial state decreases monotonically.  After we apply the dynamic-reversing operation, the fidelity starts increasing which acquire the maximum value $1$ at $2T = 194$ steps. From this plot it is clear that the system regains its initial state with high fidelity, thus, confirming the high coherence in the state.

\begin{figure}
  \includegraphics[width=0.45\textwidth]{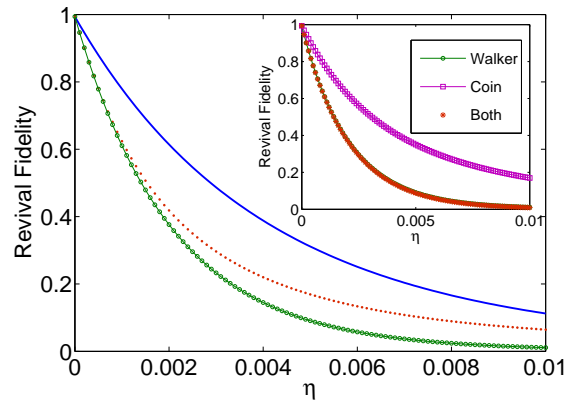}
  \caption{Revival fidelity of the quantum walk under the action of amplitude damping (blue solid curve) and bit-flip (red dotted curve) on the coin and  pure dephasing (green circles)  on the walker. The reversal operation $\sigma_y$ is applied at time $T = 250$ steps. Thus, the total evolution is for $500$ steps. In the inset we plot the revival fidelity when the pure dephasing bath is applied only on the coin (magenta squares), only on the walker (green circles) and on both (red dots). }
  \label{Fig:Decoherence}
\end{figure}

In Fig.~\ref{Fig:Decoherence} we plot the revival fidelity as a function of the bath strength $\eta$ for all three types of baths. Here we have chosen $T = 250$ steps, thus, the total evolution is for $2T = 500$ steps.
This figure shows that we can achieve a very high revival fidelity for small $\eta$ ($\eta \approx 0.001$). If we choose $T$ to be smaller then the revival fidelity can be high even for stronger bath interactions. Another important point to observe is that the cat states in the amplitude damping bath (solid blue curve) seem to do better than the bit-flip (red dotted curve) and the pure dephasing bath (green circles). In the inset we can see the effect of applying pure dephasing only on the coin, only on the walker and on both. Clearly, when the bath is acting only on the coin space the cat states can survive longer than when the bath acts on the walker or on the combined walker plus coin space. 
Overall Fig.~\ref{Fig:Decoherence} suggests that the cat states with significant separation  between the two components in the bimodal distribution should be possible in the physical implementations of quantum walks.

\section{Controlling the cat states in OAM implementation of quantum walks}\label{Sec:Control}
In this section, we consider the optical implementation of the one-dimensional quantum walk which we introduced  in Sec.~\ref{Sub:Implementation}. In this implementation, the quantum walk is performed over the OAM space of light. Here we propose a method to manipulate and control the separation between the two distinct components in the cat state.

The first requirement to realize a cat state in a one-dimensional quantum walk is the delocalized (Gaussian) initial state. The Gaussian initial state in the OAM implementation of the quantum walk can be constructed, simply, by using  a spatial light modulators (SLM)~\cite{Cardano2015}. An SLM is a device that can manipulate the transverse phase and the amplitude of a light beam. Since the OAM of light is due to the azimuthal phase profile in the transverse plane, SLM is the perfect device to achieve an arbitrary superposition of the OAM states. In order to create a delocalized initial state in the OAM basis a phase hologram is displayed on the SLM device. A plane wave is reflected from this device which acquires the transverse state corresponding to the phase hologram. 
Thus, by using a spatial light modulator and using the scheme presented in Ref.~\cite{Cardano2015,Goyal2013} we can form the cat states in the optical quantum walks.

After realizing the cat state, the next step is to control the separation between the macroscopic states of the walker. In Sec.~\ref{Subsec:DTQW} we have seen that the  application of the momentum shift operator $F_{\rm m}$ in a one-dimensional DTQW causes a periodic revival of the initial state of the walker. The walker regains its initial state after $2p$ number of steps where the number $p = 2\pi/\Phi$ is related to the parameter of the operator $F_{\rm m}$.

We use the same $F_{\rm m}$ to stabilize the cat state in the quantum walks. In order to stabilize the cat state at time $t$, first, we evolve the initial Gaussian state of the walker for time $t$  using the quantum walk propagator $Z$~\eqref{Eq:Z}. The evolved state $\Ket{\Psi(t)}$  reads
\begin{equation}
  \Ket{\Psi(t)} \approx \frac{1}{\sqrt{2}} \left(\Ket{G(-n_t,\sigma)}\otimes \ket{u_-} + \Ket{G(n_t,\sigma)}\otimes \ket{u_+} \right).
\end{equation}
At time $t$ we introduce the momentum shift operator $F_{\rm m}$ in the quantum walk with a certain value of $p$.  Due to the momentum shift operator the state of the walker starts  oscillating, recovering the state $\Ket{\Psi(t)}$ periodically after the time period $2p$. Hence, we can preserve the cat state $\Ket{\Psi(t)}$ for a long time. The only obstacle  in preserving the cat states is the decoherence.

\begin{figure}
  \includegraphics[width=0.4\textwidth]{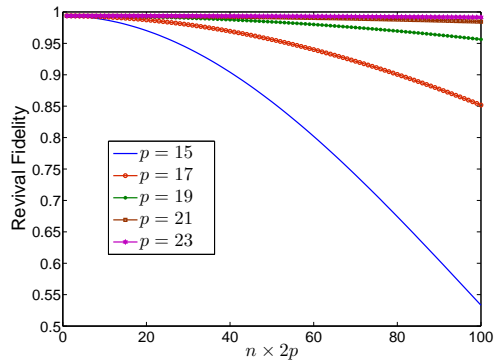}
  \caption{Here we plot the revival fidelity of the cat state after the experiencing electric field  for $n\times 2p$. Here the width of the delocalized initial state is $\sigma = 9$ and the time of evolution is $t = 100$ steps.}\label{Fig:ElectricFid}
\end{figure}

Now if we want to increase the separation between the two Gaussian wave-packets of the cat state $\Ket{\Psi(t)}$, then we remove the operator $F_{\rm m}$ after a time period which is an integer multiple of $2p$. On the other hand, if we want to decrease the separation between the two Gaussian wave-packets we remove $F_{\rm m}$ after  $2np$ followed immediately by one-time application of $\sigma_y$ operation. Hence, by introducing the $F_{\rm m}$ and the reversal operation $\sigma_y$ we can control and manipulate the cat states in the quantum walks.

In  Fig.~\ref{Fig:ElectricFid} we show the numerically calculated revival fidelity of the cat states. Here we have evolved the delocalized initial state for $100$ steps. Then we apply $F_{\rm m}$ for $n\times 2p$ number of steps. We remove the operator $F_{\rm m}$ and apply the quantum walk reversal operation $\sigma_y$ and evolve the system for $100$ steps and calculate the fidelity with the initial state. Here $n$ is an integer between $1$ and $100$. We can see that for sufficiently large values of $p$ the revival fidelity converges to the value $1$.

The action of the operator $F_{\rm m}$ can be implemented in the OAM quantum walk by means of a Dove prism~\cite{Padgett1999}. The action of the dove prism on the OAM states of light can be written as
\begin{equation}
  \label{Eq:Dove}
  \ket{\ell} \to \exp(i 2\varphi\ell) \ket{-\ell},
\end{equation}
where $\varphi$ is the angle of rotation of the dove prism along the propagation axis of the light beam. Thus, two dove prisms in a sequence with angles $\varphi/4$ and $-\varphi/4$ can implement the action of the operator $F_{\rm m}$~\eqref{Eq:MomentumShift} with $\Phi = \varphi$.

The final component required to achieve the complete control over the cat states in one-dimensional quantum walks is the reversal operation $\sigma_y$. In the current scheme, this operation can be achieved by simply using a half-wave plate that has the fast axis parallel to the horizontal axis.

Although, we can realize the $F_{\rm m}$ and the $\sigma_y$ operations using a dove prism and a half-wave plate, respectively, the real difficulty is in applying these operators at a given time $t$ and for a desired duration. To achieve this we can use optical switches in the quantum walk setup~\cite{Hall2011}. The role of an optical switch is to direct the light beam in a particular path that can be selected dynamically.  Thus, by using optical switches we can make the light beam in the OAM quantum walk setup to pass through the free space which will result in the normal quantum walk. At time $t$ we can trigger the optical switch to make the light beam pass through the dove prism which will restrict the spread of the state of the walker. We can again trigger the switch at any desired time so that the beam again pass through the free space or though the third path which contains the half-wave plate.  Hence, by using optical switchs, dove prisms and a half-wave plate we can gain total control over the cat states in one-dimensional quantum walks.

To summarize, we have discussed an optical scheme to manipulate and control the cat state in OAM quantum walks using linear optical devices  half-wave plates and dove prisms, and optical switches.

\section{Conclusion }
\label{Sec:Conclude}
In conclusion, we have proposed a method to prepare  cat states in  quantum walk setups using delocalized initial states. Our method is system-independent and works for the entire range of the parameter $\theta$.  We have also studied the effects of environmental interactions on the  cat states and demonstrated that large separation in the cat states is possible even in the presence of noise. Finally, we have presented a method to control and manipulate the cat states in optical systems.

The formation of the cat states in one-dimensional DTQW yields an interesting class of small-momentum Hamiltonians that, despite being non-linear in the momentum, possess linear dispersion relation. 
Both quantum walks and cat states have been used to describe the coherent energy transfer in photosynthesis~\cite{Klafter1980,Barvik1987,Engel2007,Lee2007,Mohseni2008,Nalbach2011}. The current proposal of preparing cat states using quantum walks threads the two concepts together, which might also contribute to a better understanding of the underlying physics of photosynthesis.

\begin{acknowledgements}
SKG and CS acknowledge the support from NSERC. BCS thanks NSERC, Alberta Innovates, and China's 1000 Talent Plan for financial support. WZ appreciates the financial support from the BUPT Excellent Ph.D.~Students Foundation (Grant No.~CX201325), the China Scholarship Council (Grant  No.~201406470022), and  NSERC.  FG acknowledges the financial support from NSFC (Grants No.~61272057 and No.~61572081).
\end{acknowledgements}

\appendix
\section{Hamiltonian for the one-dimensional discrete time quantum walk}\label{App:QWalkHamil}\label{App:Dirac}
The low-momentum expansion of the Hamiltonian $H(k)$ in~\eqref{Eq:QWalkHamil2.1} can be calculated by using the Taylor series expansion of the Hamiltonian $H(k)$ and discarding the $O(k^3)$ and higher order terms. The truncated  $2 {\rm nd}$-order Hamiltonians read
\begin{widetext}
\begin{align}
  H^{(2)} &= \begin{pmatrix}-\cos\theta \left( k\cos\theta + \frac{\pi}{2} - \frac{1}{4}\pi k^2\sin^2\theta\right) &
\begin{pmatrix}-\sin\theta \left( k\cos\theta + \frac{\pi}{2} - \frac{1}{4}\pi k^2\sin^2\theta\right)\\
-ik\sin\theta\left( k\cos\theta + \frac{\pi}{2}\right)
\end{pmatrix}\\
\begin{pmatrix}-\sin\theta \left( k\cos\theta + \frac{\pi}{2} - \frac{1}{4}\pi k^2\sin^2\theta\right)\\
 +ik\sin\theta\left( k\cos\theta + \frac{\pi}{2}\right)
\end{pmatrix}
& \cos\theta \left( k\cos\theta + \frac{\pi}{2} - \frac{1}{4}\pi k^2\sin^2\theta\right)
  \end{pmatrix}.  \label{Eq:H2Matrix}
\end{align}
\end{widetext}
The eigenvalues $E_{\pm}(k)$ of the Hamiltonian~\eqref{Eq:H2Matrix} read
\begin{align}
 E_\pm(k)&=\pm\left(k \cos \theta +\frac{\pi }{2}\right) + O(k^3),
\end{align}
and the corresponding eigenvectors read
\begin{align}
\Ket{u_-(k)}=\frac{1}{N_1}\begin{pmatrix}
 \left(-\frac{1}{2}k^2\cos\theta  + k^2 - 2ik -2\right) \cos \frac{\theta }{2}\\\sin\frac{\theta }{2}\end{pmatrix}\label{Eq:Eigenvector1},
\end{align}
\begin{align}
\Ket{u_+(k)}=\frac{1}{N_2} \begin{pmatrix}
\left(-\frac{1}{2} k^2\cos\theta -k^2 + 2ik +2\right) \sin \frac{\theta }{2}\\\cos\frac{\theta }{2}\label{Eq:Eigenvector2}
\end{pmatrix}.
\end{align}
Here, $N_1$ and $N_2$ are normalization factors which read
\begin{align}
N_1&=\sqrt{\sin^2\frac{\theta }{2}+ \left|-\frac{k^2}{2}\cos\theta+ k^2 - 2ik -2\right| ^2 \cos^2 \frac{\theta }{2}},\\
N_2&=\sqrt{\cos^2\frac{\theta }{2}+ \left|\frac{k^2}{2}\cos\theta+ k^2 - 2ik -2\right| ^2 \sin^2 \frac{\theta }{2}}.
\end{align}
With these eigenvectors and eigenvalues we can rewrite the Hamiltonian $H^{(2)}(k)$ as
\begin{equation}
  H^{(2)}(k) = E_+(k) \Ket{u_+(k)}\Bra{u_+(k)} + E_-\Ket{u_-(k)}\Bra{u_-(k)}.\label{Eq:H2}
\end{equation}

For the small values of the parameter $\theta$ and small $k$, the Hamiltonian $ H(k)$ reduces to a simpler form $H_{\rm d}$ that reads~\cite{Chandrashekar2010,Strauch2006}
\begin{align}
H_{\rm d}(k)&= -\left( k + \frac{\pi}{2}\right)\sigma_z -\theta\frac{\pi}{2} \sigma_x.
\label{Eqn-Dirac-01}
\end{align}
The Hamiltonian $ H_{\rm d}$ is linear in $k$; hence, it corresponds to a two-component Dirac Hamiltonian. From Eq.~\eqref{Eqn-Dirac-01} it is clear that the parameter $\theta$ characterizes the mass and the velocity of the walker. For small values of $\theta$ the walker behaves like a quantum relativistic particle with energy
\begin{equation}
  E_d(k) = \pm\sqrt{\left( k + \frac{\pi}{2}\right)^2 + \frac{\pi^2}{4}\theta^2} = \pm\left( k + \frac{\pi}{2}\right) + O(\theta^2).\label{Eq:DiracDispersion}
\end{equation}

The Hamiltonian $H_{\rm d}$ is valid only for the small values of the parameter $\theta$. For large values of $\theta$ (but still small $k$) the effective quantum walk Hamiltonian takes a slightly more complicated form which, is the truncated $1{\rm st}$-order Hamiltonian $H^{(1)}$ for quantum walks
\begin{widetext}
\begin{equation}
  H^{(1)} = \begin{pmatrix}
-\cos\theta\left(k\cos\theta + \frac{\pi}{2}\right) &
-\sin\theta\left(k\cos\theta + \frac{\pi}{2}\right)-ik\frac{\pi}{2}\sin\theta\\
-\sin\theta\left(k\cos\theta + \frac{\pi}{2}\right)+ik\frac{\pi}{2}\sin\theta &
\cos\theta\left(k\cos\theta + \frac{\pi}{2}\right)
\end{pmatrix},\label{Eq:H1Matrix}
\end{equation}
\end{widetext}
The eigenvalues for this Hamiltonian are
\begin{equation}
  E^{(1)}_\pm = \pm\sqrt{\left(k\cos\theta + \frac{\pi}{2}\right)^2 + \left(k\frac{\pi}{2}\sin\theta\right)^2},\label{Eq:E1}
\end{equation}
which are, in general, not linear in $k$. However, one can recover the linear dispersion relation~\eqref{Eq:DiracDispersion} from~\eqref{Eq:E1} by restricting the parameter $\theta$ to small values or introducing $O(k^2)$ terms in the Hamiltonian.



\begin{thebibliography}{71}%
\makeatletter
\providecommand \@ifxundefined [1]{%
 \@ifx{#1\undefined}
}%
\providecommand \@ifnum [1]{%
 \ifnum #1\expandafter \@firstoftwo
 \else \expandafter \@secondoftwo
 \fi
}%
\providecommand \@ifx [1]{%
 \ifx #1\expandafter \@firstoftwo
 \else \expandafter \@secondoftwo
 \fi
}%
\providecommand \natexlab [1]{#1}%
\providecommand \enquote  [1]{``#1''}%
\providecommand \bibnamefont  [1]{#1}%
\providecommand \bibfnamefont [1]{#1}%
\providecommand \citenamefont [1]{#1}%
\providecommand \href@noop [0]{\@secondoftwo}%
\providecommand \href [0]{\begingroup \@sanitize@url \@href}%
\providecommand \@href[1]{\@@startlink{#1}\@@href}%
\providecommand \@@href[1]{\endgroup#1\@@endlink}%
\providecommand \@sanitize@url [0]{\catcode `\\12\catcode `\$12\catcode
  `\&12\catcode `\#12\catcode `\^12\catcode `\_12\catcode `\%12\relax}%
\providecommand \@@startlink[1]{}%
\providecommand \@@endlink[0]{}%
\providecommand \url  [0]{\begingroup\@sanitize@url \@url }%
\providecommand \@url [1]{\endgroup\@href {#1}{\urlprefix }}%
\providecommand \urlprefix  [0]{URL }%
\providecommand \Eprint [0]{\href }%
\providecommand \doibase [0]{http://dx.doi.org/}%
\providecommand \selectlanguage [0]{\@gobble}%
\providecommand \bibinfo  [0]{\@secondoftwo}%
\providecommand \bibfield  [0]{\@secondoftwo}%
\providecommand \translation [1]{[#1]}%
\providecommand \BibitemOpen [0]{}%
\providecommand \bibitemStop [0]{}%
\providecommand \bibitemNoStop [0]{.\EOS\space}%
\providecommand \EOS [0]{\spacefactor3000\relax}%
\providecommand \BibitemShut  [1]{\csname bibitem#1\endcsname}%
\let\auto@bib@innerbib\@empty
\bibitem [{\citenamefont {Schr\"odinger}(1935)}]{Schrodinger1935}%
  \BibitemOpen
  \bibfield  {author} {\bibinfo {author} {\bibfnamefont {E.}~\bibnamefont
  {Schr\"odinger}},\ }\href@noop {} {\bibfield  {journal} {\bibinfo  {journal}
  {Naturwissenschaften}\ }\textbf {\bibinfo {volume} {23}},\ \bibinfo {pages}
  {807 } (\bibinfo {year} {1935})}\BibitemShut {NoStop}%
\bibitem [{\citenamefont {Milburn}(1986)}]{Milburn1986a}%
  \BibitemOpen
  \bibfield  {author} {\bibinfo {author} {\bibfnamefont {G.~J.}\ \bibnamefont
  {Milburn}},\ }\href@noop {} {\bibfield  {journal} {\bibinfo  {journal} {Phys.
  Rev. A}\ }\textbf {\bibinfo {volume} {33}},\ \bibinfo {pages} {674 }
  (\bibinfo {year} {1986})}\BibitemShut {NoStop}%
\bibitem [{\citenamefont {Yurke}\ and\ \citenamefont
  {Stoler}(1986)}]{Yurke1986}%
  \BibitemOpen
  \bibfield  {author} {\bibinfo {author} {\bibfnamefont {B.}~\bibnamefont
  {Yurke}}\ and\ \bibinfo {author} {\bibfnamefont {D.}~\bibnamefont {Stoler}},\
  }\href {\doibase 10.1103/physrevlett.57.13} {\bibfield  {journal} {\bibinfo
  {journal} {Phys. Rev. Lett.}\ }\textbf {\bibinfo {volume} {57}},\ \bibinfo
  {pages} {13 } (\bibinfo {year} {1986})}\BibitemShut {NoStop}%
\bibitem [{\citenamefont {Bu\v{z}ek}\ and\ \citenamefont
  {Knight}(1995)}]{Buzek1995}%
  \BibitemOpen
  \bibfield  {author} {\bibinfo {author} {\bibfnamefont {V.}~\bibnamefont
  {Bu\v{z}ek}}\ and\ \bibinfo {author} {\bibfnamefont {P.~L.}\ \bibnamefont
  {Knight}},\ }\href {\doibase 10.1016/s0079-6638(08)70324-x} {\bibfield
  {journal} {\bibinfo  {journal} {Prog. Opt.}\ }\textbf {\bibinfo {volume}
  {34}},\ \bibinfo {pages} {1 } (\bibinfo {year} {1995})}\BibitemShut {NoStop}%
\bibitem [{\citenamefont {Sanders}(1992)}]{Sanders1992}%
  \BibitemOpen
  \bibfield  {author} {\bibinfo {author} {\bibfnamefont {B.~C.}\ \bibnamefont
  {Sanders}},\ }\href@noop {} {\bibfield  {journal} {\bibinfo  {journal} {Phys.
  Rev. A}\ }\textbf {\bibinfo {volume} {45}},\ \bibinfo {pages} {6811 }
  (\bibinfo {year} {1992})}\BibitemShut {NoStop}%
\bibitem [{\citenamefont {Munro}\ \emph {et~al.}(2002)\citenamefont {Munro},
  \citenamefont {Nemoto}, \citenamefont {Milburn},\ and\ \citenamefont
  {Braunstein}}]{Munro2002}%
  \BibitemOpen
  \bibfield  {author} {\bibinfo {author} {\bibfnamefont {W.~J.}\ \bibnamefont
  {Munro}}, \bibinfo {author} {\bibfnamefont {K.}~\bibnamefont {Nemoto}},
  \bibinfo {author} {\bibfnamefont {G.~J.}\ \bibnamefont {Milburn}}, \ and\
  \bibinfo {author} {\bibfnamefont {S.~L.}\ \bibnamefont {Braunstein}},\
  }\href@noop {} {\bibfield  {journal} {\bibinfo  {journal} {Phys. Rev. A}\
  }\textbf {\bibinfo {volume} {66}},\ \bibinfo {pages} {023819} (\bibinfo
  {year} {2002})}\BibitemShut {NoStop}%
\bibitem [{\citenamefont {Wenger}\ \emph {et~al.}(2003)\citenamefont {Wenger},
  \citenamefont {Hafezi}, \citenamefont {Grosshans}, \citenamefont
  {Tualle-Brouri},\ and\ \citenamefont {Grangier}}]{Wenger2003}%
  \BibitemOpen
  \bibfield  {author} {\bibinfo {author} {\bibfnamefont {J.}~\bibnamefont
  {Wenger}}, \bibinfo {author} {\bibfnamefont {M.}~\bibnamefont {Hafezi}},
  \bibinfo {author} {\bibfnamefont {F.}~\bibnamefont {Grosshans}}, \bibinfo
  {author} {\bibfnamefont {R.}~\bibnamefont {Tualle-Brouri}}, \ and\ \bibinfo
  {author} {\bibfnamefont {P.}~\bibnamefont {Grangier}},\ }\href {\doibase
  10.1103/physreva.67.012105} {\bibfield  {journal} {\bibinfo  {journal} {Phys.
  Rev. A}\ }\textbf {\bibinfo {volume} {67}},\ \bibinfo {pages} {012105}
  (\bibinfo {year} {2003})}\BibitemShut {NoStop}%
\bibitem [{\citenamefont {Stobi\'nska}\ \emph {et~al.}(2007)\citenamefont
  {Stobi\'nska}, \citenamefont {Jeong},\ and\ \citenamefont
  {Ralph}}]{Stobinska2007}%
  \BibitemOpen
  \bibfield  {author} {\bibinfo {author} {\bibfnamefont {M.}~\bibnamefont
  {Stobi\'nska}}, \bibinfo {author} {\bibfnamefont {H.}~\bibnamefont {Jeong}},
  \ and\ \bibinfo {author} {\bibfnamefont {T.~C.}\ \bibnamefont {Ralph}},\
  }\href {\doibase 10.1103/physreva.75.052105} {\bibfield  {journal} {\bibinfo
  {journal} {Phys. Rev. A}\ }\textbf {\bibinfo {volume} {75}},\ \bibinfo
  {pages} {052105} (\bibinfo {year} {2007})}\BibitemShut {NoStop}%
\bibitem [{\citenamefont {Monroe}\ \emph {et~al.}(1996)\citenamefont {Monroe},
  \citenamefont {Meekhof}, \citenamefont {King},\ and\ \citenamefont
  {Wineland}}]{Monroe1996}%
  \BibitemOpen
  \bibfield  {author} {\bibinfo {author} {\bibfnamefont {C.}~\bibnamefont
  {Monroe}}, \bibinfo {author} {\bibfnamefont {D.~M.}\ \bibnamefont {Meekhof}},
  \bibinfo {author} {\bibfnamefont {B.~E.}\ \bibnamefont {King}}, \ and\
  \bibinfo {author} {\bibfnamefont {D.~J.}\ \bibnamefont {Wineland}},\ }\href
  {\doibase 10.1126/science.272.5265.1131} {\bibfield  {journal} {\bibinfo
  {journal} {Science}\ }\textbf {\bibinfo {volume} {272}},\ \bibinfo {pages}
  {1131 } (\bibinfo {year} {1996})}\BibitemShut {NoStop}%
\bibitem [{\citenamefont {Brune}\ \emph {et~al.}(1996)\citenamefont {Brune},
  \citenamefont {Hagley}, \citenamefont {Dreyer}, \citenamefont {Ma\^{i}tre},
  \citenamefont {Maali}, \citenamefont {Wunderlich}, \citenamefont {Raimond},\
  and\ \citenamefont {Haroche}}]{Brune1996}%
  \BibitemOpen
  \bibfield  {author} {\bibinfo {author} {\bibfnamefont {M.}~\bibnamefont
  {Brune}}, \bibinfo {author} {\bibfnamefont {E.}~\bibnamefont {Hagley}},
  \bibinfo {author} {\bibfnamefont {J.}~\bibnamefont {Dreyer}}, \bibinfo
  {author} {\bibfnamefont {X.}~\bibnamefont {Ma\^{i}tre}}, \bibinfo {author}
  {\bibfnamefont {A.}~\bibnamefont {Maali}}, \bibinfo {author} {\bibfnamefont
  {C.}~\bibnamefont {Wunderlich}}, \bibinfo {author} {\bibfnamefont {J.~M.}\
  \bibnamefont {Raimond}}, \ and\ \bibinfo {author} {\bibfnamefont
  {S.}~\bibnamefont {Haroche}},\ }\href {\doibase 10.1103/physrevlett.77.4887}
  {\bibfield  {journal} {\bibinfo  {journal} {Phys. Rev. Lett.}\ }\textbf
  {\bibinfo {volume} {77}},\ \bibinfo {pages} {4887 } (\bibinfo {year}
  {1996})}\BibitemShut {NoStop}%
\bibitem [{\citenamefont {Slosser}\ and\ \citenamefont
  {Milburn}(1995)}]{Slosser1995}%
  \BibitemOpen
  \bibfield  {author} {\bibinfo {author} {\bibfnamefont {J.~J.}\ \bibnamefont
  {Slosser}}\ and\ \bibinfo {author} {\bibfnamefont {G.~J.}\ \bibnamefont
  {Milburn}},\ }\href {\doibase 10.1103/physrevlett.75.418} {\bibfield
  {journal} {\bibinfo  {journal} {Phys. Rev. Lett.}\ }\textbf {\bibinfo
  {volume} {75}},\ \bibinfo {pages} {418 } (\bibinfo {year}
  {1995})}\BibitemShut {NoStop}%
\bibitem [{\citenamefont {Jeong}\ \emph {et~al.}(2004)\citenamefont {Jeong},
  \citenamefont {Kim}, \citenamefont {Ralph},\ and\ \citenamefont
  {Ham}}]{Jeong2004}%
  \BibitemOpen
  \bibfield  {author} {\bibinfo {author} {\bibfnamefont {H.}~\bibnamefont
  {Jeong}}, \bibinfo {author} {\bibfnamefont {M.~S.}\ \bibnamefont {Kim}},
  \bibinfo {author} {\bibfnamefont {T.~C.}\ \bibnamefont {Ralph}}, \ and\
  \bibinfo {author} {\bibfnamefont {B.~S.}\ \bibnamefont {Ham}},\ }\href@noop
  {} {\bibfield  {journal} {\bibinfo  {journal} {Phys. Rev. A}\ }\textbf
  {\bibinfo {volume} {70}},\ \bibinfo {pages} {061801} (\bibinfo {year}
  {2004})}\BibitemShut {NoStop}%
\bibitem [{\citenamefont {Huang}\ and\ \citenamefont
  {Moore}(2006)}]{Huang2006}%
  \BibitemOpen
  \bibfield  {author} {\bibinfo {author} {\bibfnamefont {Y.~P.}\ \bibnamefont
  {Huang}}\ and\ \bibinfo {author} {\bibfnamefont {M.~G.}\ \bibnamefont
  {Moore}},\ }\href@noop {} {\bibfield  {journal} {\bibinfo  {journal} {Phys.
  Rev. A}\ }\textbf {\bibinfo {volume} {73}},\ \bibinfo {pages} {023606}
  (\bibinfo {year} {2006})}\BibitemShut {NoStop}%
\bibitem [{\citenamefont {Bhaktavatsala~Rao}\ \emph {et~al.}(2011)\citenamefont
  {Bhaktavatsala~Rao}, \citenamefont {Bar-Gill},\ and\ \citenamefont
  {Kurizki}}]{Rao2011}%
  \BibitemOpen
  \bibfield  {author} {\bibinfo {author} {\bibfnamefont {D.~D.}\ \bibnamefont
  {Bhaktavatsala~Rao}}, \bibinfo {author} {\bibfnamefont {N.}~\bibnamefont
  {Bar-Gill}}, \ and\ \bibinfo {author} {\bibfnamefont {G.}~\bibnamefont
  {Kurizki}},\ }\href@noop {} {\bibfield  {journal} {\bibinfo  {journal} {Phys.
  Rev. Lett.}\ }\textbf {\bibinfo {volume} {106}},\ \bibinfo {pages} {010404}
  (\bibinfo {year} {2011})}\BibitemShut {NoStop}%
\bibitem [{\citenamefont {Csire}\ and\ \citenamefont
  {Apagyi}(2012)}]{Csire2012}%
  \BibitemOpen
  \bibfield  {author} {\bibinfo {author} {\bibfnamefont {G.}~\bibnamefont
  {Csire}}\ and\ \bibinfo {author} {\bibfnamefont {B.}~\bibnamefont {Apagyi}},\
  }\href@noop {} {\bibfield  {journal} {\bibinfo  {journal} {Phys. Rev. A}\
  }\textbf {\bibinfo {volume} {85}},\ \bibinfo {pages} {033613} (\bibinfo
  {year} {2012})}\BibitemShut {NoStop}%
\bibitem [{\citenamefont {Lee}\ \emph {et~al.}(2012)\citenamefont {Lee},
  \citenamefont {Lee}, \citenamefont {Nha},\ and\ \citenamefont
  {Jeong}}]{Lee2012}%
  \BibitemOpen
  \bibfield  {author} {\bibinfo {author} {\bibfnamefont {C.-W.}\ \bibnamefont
  {Lee}}, \bibinfo {author} {\bibfnamefont {J.}~\bibnamefont {Lee}}, \bibinfo
  {author} {\bibfnamefont {H.}~\bibnamefont {Nha}}, \ and\ \bibinfo {author}
  {\bibfnamefont {H.}~\bibnamefont {Jeong}},\ }\href@noop {} {\bibfield
  {journal} {\bibinfo  {journal} {Phys. Rev. A}\ }\textbf {\bibinfo {volume}
  {85}},\ \bibinfo {pages} {063815} (\bibinfo {year} {2012})}\BibitemShut
  {NoStop}%
\bibitem [{\citenamefont {Wu}\ and\ \citenamefont {Zhang}(2013)}]{Wu2013}%
  \BibitemOpen
  \bibfield  {author} {\bibinfo {author} {\bibfnamefont {B.}~\bibnamefont
  {Wu}}\ and\ \bibinfo {author} {\bibfnamefont {J.}~\bibnamefont {Zhang}},\
  }\href {\doibase 10.1007/s11433-013-5152-z} {\bibfield  {journal} {\bibinfo
  {journal} {Sci. China Phys. Mech. Astron.}\ }\textbf {\bibinfo {volume}
  {56}},\ \bibinfo {pages} {1810 } (\bibinfo {year} {2013})}\BibitemShut
  {NoStop}%
\bibitem [{\citenamefont {Lau}\ \emph {et~al.}(2014)\citenamefont {Lau},
  \citenamefont {Dutton}, \citenamefont {Wang},\ and\ \citenamefont
  {Simon}}]{Lau2014}%
  \BibitemOpen
  \bibfield  {author} {\bibinfo {author} {\bibfnamefont {H.~W.}\ \bibnamefont
  {Lau}}, \bibinfo {author} {\bibfnamefont {Z.}~\bibnamefont {Dutton}},
  \bibinfo {author} {\bibfnamefont {T.}~\bibnamefont {Wang}}, \ and\ \bibinfo
  {author} {\bibfnamefont {C.}~\bibnamefont {Simon}},\ }\href@noop {}
  {\bibfield  {journal} {\bibinfo  {journal} {Phys. Rev. Lett.}\ }\textbf
  {\bibinfo {volume} {113}},\ \bibinfo {pages} {090401} (\bibinfo {year}
  {2014})}\BibitemShut {NoStop}%
\bibitem [{\citenamefont {Fischer}\ and\ \citenamefont
  {Kang}(2015)}]{Fischer2015}%
  \BibitemOpen
  \bibfield  {author} {\bibinfo {author} {\bibfnamefont {U.~R.}\ \bibnamefont
  {Fischer}}\ and\ \bibinfo {author} {\bibfnamefont {M.-K.}\ \bibnamefont
  {Kang}},\ }\href@noop {} {\bibfield  {journal} {\bibinfo  {journal} {Phys.
  Rev. Lett.}\ }\textbf {\bibinfo {volume} {115}},\ \bibinfo {pages} {260404}
  (\bibinfo {year} {2015})}\BibitemShut {NoStop}%
\bibitem [{\citenamefont {Wang}\ \emph {et~al.}(2015)\citenamefont {Wang},
  \citenamefont {Lau}, \citenamefont {Kaviani}, \citenamefont {Ghobadi},\ and\
  \citenamefont {Simon}}]{Wang2015}%
  \BibitemOpen
  \bibfield  {author} {\bibinfo {author} {\bibfnamefont {T.}~\bibnamefont
  {Wang}}, \bibinfo {author} {\bibfnamefont {H.~W.}\ \bibnamefont {Lau}},
  \bibinfo {author} {\bibfnamefont {H.}~\bibnamefont {Kaviani}}, \bibinfo
  {author} {\bibfnamefont {R.}~\bibnamefont {Ghobadi}}, \ and\ \bibinfo
  {author} {\bibfnamefont {C.}~\bibnamefont {Simon}},\ }\href {\doibase
  10.1103/physreva.92.012316} {\bibfield  {journal} {\bibinfo  {journal} {Phys.
  Rev. A}\ }\textbf {\bibinfo {volume} {92}},\ \bibinfo {pages} {012316}
  (\bibinfo {year} {2015})}\BibitemShut {NoStop}%
\bibitem [{\citenamefont {Friedman}\ \emph {et~al.}(2000)\citenamefont
  {Friedman}, \citenamefont {Patel}, \citenamefont {Chen}, \citenamefont
  {Tolpygo},\ and\ \citenamefont {Lukens}}]{Friedman2000}%
  \BibitemOpen
  \bibfield  {author} {\bibinfo {author} {\bibfnamefont {J.~R.}\ \bibnamefont
  {Friedman}}, \bibinfo {author} {\bibfnamefont {V.}~\bibnamefont {Patel}},
  \bibinfo {author} {\bibfnamefont {W.}~\bibnamefont {Chen}}, \bibinfo {author}
  {\bibfnamefont {S.~K.}\ \bibnamefont {Tolpygo}}, \ and\ \bibinfo {author}
  {\bibfnamefont {J.~E.}\ \bibnamefont {Lukens}},\ }\href@noop {} {\bibfield
  {journal} {\bibinfo  {journal} {Nature}\ }\textbf {\bibinfo {volume} {406}},\
  \bibinfo {pages} {43 } (\bibinfo {year} {2000})}\BibitemShut {NoStop}%
\bibitem [{\citenamefont {Leibfried}\ \emph {et~al.}(2005)\citenamefont
  {Leibfried}, \citenamefont {Knill}, \citenamefont {Seidelin}, \citenamefont
  {Britton}, \citenamefont {Blakestad}, \citenamefont {Chiaverini},
  \citenamefont {Hume}, \citenamefont {Itano}, \citenamefont {Jost},
  \citenamefont {Langer}, \citenamefont {Ozeri1}, \citenamefont {Reichle1},\
  and\ \citenamefont {Wineland}}]{Leibfried2005}%
  \BibitemOpen
  \bibfield  {author} {\bibinfo {author} {\bibfnamefont {D.}~\bibnamefont
  {Leibfried}}, \bibinfo {author} {\bibfnamefont {E.}~\bibnamefont {Knill}},
  \bibinfo {author} {\bibfnamefont {S.}~\bibnamefont {Seidelin}}, \bibinfo
  {author} {\bibfnamefont {J.}~\bibnamefont {Britton}}, \bibinfo {author}
  {\bibfnamefont {R.~B.}\ \bibnamefont {Blakestad}}, \bibinfo {author}
  {\bibfnamefont {J.}~\bibnamefont {Chiaverini}}, \bibinfo {author}
  {\bibfnamefont {D.~B.}\ \bibnamefont {Hume}}, \bibinfo {author}
  {\bibfnamefont {W.~M.}\ \bibnamefont {Itano}}, \bibinfo {author}
  {\bibfnamefont {J.~D.}\ \bibnamefont {Jost}}, \bibinfo {author}
  {\bibfnamefont {C.}~\bibnamefont {Langer}}, \bibinfo {author} {\bibfnamefont
  {R.}~\bibnamefont {Ozeri1}}, \bibinfo {author} {\bibfnamefont
  {R.}~\bibnamefont {Reichle1}}, \ and\ \bibinfo {author} {\bibfnamefont
  {D.~J.}\ \bibnamefont {Wineland}},\ }\href@noop {} {\bibfield  {journal}
  {\bibinfo  {journal} {Nature}\ }\textbf {\bibinfo {volume} {438}},\ \bibinfo
  {pages} {639 } (\bibinfo {year} {2005})}\BibitemShut {NoStop}%
\bibitem [{\citenamefont {Ourjoumtsev}\ \emph {et~al.}(2007)\citenamefont
  {Ourjoumtsev}, \citenamefont {Jeong}, \citenamefont {Tualle-Brouri},\ and\
  \citenamefont {Grangier}}]{Ourjoumtsev2007}%
  \BibitemOpen
  \bibfield  {author} {\bibinfo {author} {\bibfnamefont {A.}~\bibnamefont
  {Ourjoumtsev}}, \bibinfo {author} {\bibfnamefont {H.}~\bibnamefont {Jeong}},
  \bibinfo {author} {\bibfnamefont {R.}~\bibnamefont {Tualle-Brouri}}, \ and\
  \bibinfo {author} {\bibfnamefont {P.}~\bibnamefont {Grangier}},\ }\href@noop
  {} {\bibfield  {journal} {\bibinfo  {journal} {Nature}\ }\textbf {\bibinfo
  {volume} {448}},\ \bibinfo {pages} {784 } (\bibinfo {year}
  {2007})}\BibitemShut {NoStop}%
\bibitem [{\citenamefont {Lvovsky}\ \emph {et~al.}(2013)\citenamefont
  {Lvovsky}, \citenamefont {Ghobadi}, \citenamefont {Chandra}, \citenamefont
  {Prasad},\ and\ \citenamefont {Simon}}]{Lvovsky2013}%
  \BibitemOpen
  \bibfield  {author} {\bibinfo {author} {\bibfnamefont {A.~I.}\ \bibnamefont
  {Lvovsky}}, \bibinfo {author} {\bibfnamefont {R.}~\bibnamefont {Ghobadi}},
  \bibinfo {author} {\bibfnamefont {A.}~\bibnamefont {Chandra}}, \bibinfo
  {author} {\bibfnamefont {A.~S.}\ \bibnamefont {Prasad}}, \ and\ \bibinfo
  {author} {\bibfnamefont {C.}~\bibnamefont {Simon}},\ }\href@noop {}
  {\bibfield  {journal} {\bibinfo  {journal} {Nat. Phys.}\ }\textbf {\bibinfo
  {volume} {9}},\ \bibinfo {pages} {541 } (\bibinfo {year} {2013})}\BibitemShut
  {NoStop}%
\bibitem [{\citenamefont {Bruno}\ \emph {et~al.}(2013)\citenamefont {Bruno},
  \citenamefont {Martin}, \citenamefont {Sekatski}, \citenamefont {Sangouard},
  \citenamefont {Thew},\ and\ \citenamefont {Gisin}}]{Bruno2013}%
  \BibitemOpen
  \bibfield  {author} {\bibinfo {author} {\bibfnamefont {N.}~\bibnamefont
  {Bruno}}, \bibinfo {author} {\bibfnamefont {A.}~\bibnamefont {Martin}},
  \bibinfo {author} {\bibfnamefont {P.}~\bibnamefont {Sekatski}}, \bibinfo
  {author} {\bibfnamefont {N.}~\bibnamefont {Sangouard}}, \bibinfo {author}
  {\bibfnamefont {R.~T.}\ \bibnamefont {Thew}}, \ and\ \bibinfo {author}
  {\bibfnamefont {N.}~\bibnamefont {Gisin}},\ }\href@noop {} {\bibfield
  {journal} {\bibinfo  {journal} {Nat. Phys.}\ }\textbf {\bibinfo {volume}
  {9}},\ \bibinfo {pages} {545 } (\bibinfo {year} {2013})}\BibitemShut
  {NoStop}%
\bibitem [{\citenamefont {Vlastakis}\ \emph {et~al.}(2013)\citenamefont
  {Vlastakis}, \citenamefont {Kirchmair}, \citenamefont {Leghtas},
  \citenamefont {Nigg}, \citenamefont {Frunzio}, \citenamefont {Girvin},
  \citenamefont {Mirrahimi}, \citenamefont {Devoret},\ and\ \citenamefont
  {Schoelkopf}}]{Vlastakis2013}%
  \BibitemOpen
  \bibfield  {author} {\bibinfo {author} {\bibfnamefont {B.}~\bibnamefont
  {Vlastakis}}, \bibinfo {author} {\bibfnamefont {G.}~\bibnamefont
  {Kirchmair}}, \bibinfo {author} {\bibfnamefont {Z.}~\bibnamefont {Leghtas}},
  \bibinfo {author} {\bibfnamefont {S.~E.}\ \bibnamefont {Nigg}}, \bibinfo
  {author} {\bibfnamefont {L.}~\bibnamefont {Frunzio}}, \bibinfo {author}
  {\bibfnamefont {S.~M.}\ \bibnamefont {Girvin}}, \bibinfo {author}
  {\bibfnamefont {M.}~\bibnamefont {Mirrahimi}}, \bibinfo {author}
  {\bibfnamefont {M.~H.}\ \bibnamefont {Devoret}}, \ and\ \bibinfo {author}
  {\bibfnamefont {R.~J.}\ \bibnamefont {Schoelkopf}},\ }\href {\doibase
  10.1126/science.1243289} {\bibfield  {journal} {\bibinfo  {journal}
  {Science}\ }\textbf {\bibinfo {volume} {342}},\ \bibinfo {pages} {607 }
  (\bibinfo {year} {2013})}\BibitemShut {NoStop}%
\bibitem [{\citenamefont {Wang}\ \emph {et~al.}(2016)\citenamefont {Wang},
  \citenamefont {Gao}, \citenamefont {Reinhold}, \citenamefont {Heeres},
  \citenamefont {Ofek}, \citenamefont {Chou}, \citenamefont {Axline},
  \citenamefont {Reagor}, \citenamefont {Blumoff}, \citenamefont {Sliwa},\ and\
  \citenamefont {et~al.}}]{Wang2016}%
  \BibitemOpen
  \bibfield  {author} {\bibinfo {author} {\bibfnamefont {C.}~\bibnamefont
  {Wang}}, \bibinfo {author} {\bibfnamefont {Y.~Y.}\ \bibnamefont {Gao}},
  \bibinfo {author} {\bibfnamefont {P.}~\bibnamefont {Reinhold}}, \bibinfo
  {author} {\bibfnamefont {R.~W.}\ \bibnamefont {Heeres}}, \bibinfo {author}
  {\bibfnamefont {N.}~\bibnamefont {Ofek}}, \bibinfo {author} {\bibfnamefont
  {K.}~\bibnamefont {Chou}}, \bibinfo {author} {\bibfnamefont {C.}~\bibnamefont
  {Axline}}, \bibinfo {author} {\bibfnamefont {M.}~\bibnamefont {Reagor}},
  \bibinfo {author} {\bibfnamefont {J.}~\bibnamefont {Blumoff}}, \bibinfo
  {author} {\bibfnamefont {K.~M.}\ \bibnamefont {Sliwa}}, \ and\ \bibinfo
  {author} {\bibnamefont {et~al.}},\ }\href {\doibase 10.1126/science.aaf2941}
  {\bibfield  {journal} {\bibinfo  {journal} {Science}\ }\textbf {\bibinfo
  {volume} {352}},\ \bibinfo {pages} {1087 } (\bibinfo {year}
  {2016})}\BibitemShut {NoStop}%
\bibitem [{\citenamefont {Venegas-Andraca}(2012)}]{Venegas-Andraca2012}%
  \BibitemOpen
  \bibfield  {author} {\bibinfo {author} {\bibfnamefont {S.~E.}\ \bibnamefont
  {Venegas-Andraca}},\ }\href {\doibase 10.1007/s11128-012-0432-5} {\bibfield
  {journal} {\bibinfo  {journal} {Quantum Inf. Process.}\ }\textbf {\bibinfo
  {volume} {11}},\ \bibinfo {pages} {1015 } (\bibinfo {year}
  {2012})}\BibitemShut {NoStop}%
\bibitem [{\citenamefont {Kempe}(2003)}]{Kempe2003}%
  \BibitemOpen
  \bibfield  {author} {\bibinfo {author} {\bibfnamefont {J.}~\bibnamefont
  {Kempe}},\ }\href {\doibase 10.1080/00107151031000110776} {\bibfield
  {journal} {\bibinfo  {journal} {Contemp. Phys.}\ }\textbf {\bibinfo {volume}
  {44}},\ \bibinfo {pages} {307 } (\bibinfo {year} {2003})}\BibitemShut
  {NoStop}%
\bibitem [{\citenamefont {Ambainis}(2003)}]{Ambainis2003}%
  \BibitemOpen
  \bibfield  {author} {\bibinfo {author} {\bibfnamefont {A.}~\bibnamefont
  {Ambainis}},\ }\href@noop {} {\bibfield  {journal} {\bibinfo  {journal} {Int.
  J. Quantum Inf.}\ }\textbf {\bibinfo {volume} {01}},\ \bibinfo {pages}
  {507–518} (\bibinfo {year} {2003})}\BibitemShut {NoStop}%
\bibitem [{\citenamefont {Shenvi}\ \emph {et~al.}(2003)\citenamefont {Shenvi},
  \citenamefont {Kempe},\ and\ \citenamefont {Whaley}}]{Shenvi2003}%
  \BibitemOpen
  \bibfield  {author} {\bibinfo {author} {\bibfnamefont {N.}~\bibnamefont
  {Shenvi}}, \bibinfo {author} {\bibfnamefont {J.}~\bibnamefont {Kempe}}, \
  and\ \bibinfo {author} {\bibfnamefont {K.~B.}\ \bibnamefont {Whaley}},\
  }\href {\doibase 10.1103/physreva.67.052307} {\bibfield  {journal} {\bibinfo
  {journal} {Phys. Rev. A}\ }\textbf {\bibinfo {volume} {67}},\ \bibinfo
  {pages} {052307} (\bibinfo {year} {2003})}\BibitemShut {NoStop}%
\bibitem [{\citenamefont {Childs}\ and\ \citenamefont
  {Goldstone}(2004)}]{Childs2004}%
  \BibitemOpen
  \bibfield  {author} {\bibinfo {author} {\bibfnamefont {A.~M.}\ \bibnamefont
  {Childs}}\ and\ \bibinfo {author} {\bibfnamefont {J.}~\bibnamefont
  {Goldstone}},\ }\href@noop {} {\bibfield  {journal} {\bibinfo  {journal}
  {Phys. Rev. A}\ }\textbf {\bibinfo {volume} {70}},\ \bibinfo {pages} {022314}
  (\bibinfo {year} {2004})}\BibitemShut {NoStop}%
\bibitem [{\citenamefont {Childs}(2009{\natexlab{a}})}]{Childs2009a}%
  \BibitemOpen
  \bibfield  {author} {\bibinfo {author} {\bibfnamefont {A.~M.}\ \bibnamefont
  {Childs}},\ }\href@noop {} {\bibfield  {journal} {\bibinfo  {journal} {Phys.
  Rev. Lett.}\ }\textbf {\bibinfo {volume} {102}},\ \bibinfo {pages} {180501}
  (\bibinfo {year} {2009}{\natexlab{a}})}\BibitemShut {NoStop}%
\bibitem [{\citenamefont {Lovett}\ \emph {et~al.}(2010)\citenamefont {Lovett},
  \citenamefont {Cooper}, \citenamefont {Everitt}, \citenamefont {Trevers},\
  and\ \citenamefont {Kendon}}]{Lovett2010}%
  \BibitemOpen
  \bibfield  {author} {\bibinfo {author} {\bibfnamefont {N.~B.}\ \bibnamefont
  {Lovett}}, \bibinfo {author} {\bibfnamefont {S.}~\bibnamefont {Cooper}},
  \bibinfo {author} {\bibfnamefont {M.}~\bibnamefont {Everitt}}, \bibinfo
  {author} {\bibfnamefont {M.}~\bibnamefont {Trevers}}, \ and\ \bibinfo
  {author} {\bibfnamefont {V.}~\bibnamefont {Kendon}},\ }\href@noop {}
  {\bibfield  {journal} {\bibinfo  {journal} {Phys. Rev. A}\ }\textbf {\bibinfo
  {volume} {81}},\ \bibinfo {pages} {042330} (\bibinfo {year}
  {2010})}\BibitemShut {NoStop}%
\bibitem [{\citenamefont {Klafter}\ and\ \citenamefont
  {Silbey}(1980)}]{Klafter1980}%
  \BibitemOpen
  \bibfield  {author} {\bibinfo {author} {\bibfnamefont {J.}~\bibnamefont
  {Klafter}}\ and\ \bibinfo {author} {\bibfnamefont {R.}~\bibnamefont
  {Silbey}},\ }\href@noop {} {\bibfield  {journal} {\bibinfo  {journal} {Phys.
  Lett.}\ }\textbf {\bibinfo {volume} {125}},\ \bibinfo {pages} {339 }
  (\bibinfo {year} {1980})}\BibitemShut {NoStop}%
\bibitem [{\citenamefont {Barv\'ik}\ and\ \citenamefont
  {Sz\"ocs}(1987)}]{Barvik1987}%
  \BibitemOpen
  \bibfield  {author} {\bibinfo {author} {\bibfnamefont {I.}~\bibnamefont
  {Barv\'ik}}\ and\ \bibinfo {author} {\bibfnamefont {V.}~\bibnamefont
  {Sz\"ocs}},\ }\href@noop {} {\bibfield  {journal} {\bibinfo  {journal} {Phys.
  Lett. A}\ }\textbf {\bibinfo {volume} {125}},\ \bibinfo {pages} {339 }
  (\bibinfo {year} {1987})}\BibitemShut {NoStop}%
\bibitem [{\citenamefont {Strauch}(2006)}]{Strauch2006}%
  \BibitemOpen
  \bibfield  {author} {\bibinfo {author} {\bibfnamefont {F.~W.}\ \bibnamefont
  {Strauch}},\ }\href {\doibase 10.1103/PhysRevA.73.054302} {\bibfield
  {journal} {\bibinfo  {journal} {Phys. Rev. A}\ }\textbf {\bibinfo {volume}
  {73}},\ \bibinfo {pages} {054302} (\bibinfo {year} {2006})}\BibitemShut
  {NoStop}%
\bibitem [{\citenamefont {Strauch}(2007)}]{Strauch2007}%
  \BibitemOpen
  \bibfield  {author} {\bibinfo {author} {\bibfnamefont {F.~W.}\ \bibnamefont
  {Strauch}},\ }\href {\doibase http://dx.doi.org/10.1063/1.2759837} {\bibfield
   {journal} {\bibinfo  {journal} {J. Math. Phys.}\ }\textbf {\bibinfo {volume}
  {48}},\ \bibinfo {eid} {082102} (\bibinfo {year} {2007})}\BibitemShut
  {NoStop}%
\bibitem [{\citenamefont {Bracken}\ \emph {et~al.}(2007)\citenamefont
  {Bracken}, \citenamefont {Ellinas},\ and\ \citenamefont
  {Smyrnakis}}]{Bracken2007}%
  \BibitemOpen
  \bibfield  {author} {\bibinfo {author} {\bibfnamefont {A.~J.}\ \bibnamefont
  {Bracken}}, \bibinfo {author} {\bibfnamefont {D.}~\bibnamefont {Ellinas}}, \
  and\ \bibinfo {author} {\bibfnamefont {I.}~\bibnamefont {Smyrnakis}},\
  }\href@noop {} {\bibfield  {journal} {\bibinfo  {journal} {Phys. Rev. A}\
  }\textbf {\bibinfo {volume} {75}},\ \bibinfo {pages} {022322} (\bibinfo
  {year} {2007})}\BibitemShut {NoStop}%
\bibitem [{\citenamefont {Engel}\ \emph {et~al.}(2007)\citenamefont {Engel},
  \citenamefont {Calhoun}, \citenamefont {Read}, \citenamefont {Ahn},
  \citenamefont {Man\v{c}al}, \citenamefont {Cheng}, \citenamefont
  {Blankenship},\ and\ \citenamefont {Fleming}}]{Engel2007}%
  \BibitemOpen
  \bibfield  {author} {\bibinfo {author} {\bibfnamefont {G.~S.}\ \bibnamefont
  {Engel}}, \bibinfo {author} {\bibfnamefont {T.~R.}\ \bibnamefont {Calhoun}},
  \bibinfo {author} {\bibfnamefont {E.~L.}\ \bibnamefont {Read}}, \bibinfo
  {author} {\bibfnamefont {T.-K.}\ \bibnamefont {Ahn}}, \bibinfo {author}
  {\bibfnamefont {T.}~\bibnamefont {Man\v{c}al}}, \bibinfo {author}
  {\bibfnamefont {Y.-C.}\ \bibnamefont {Cheng}}, \bibinfo {author}
  {\bibfnamefont {R.~E.}\ \bibnamefont {Blankenship}}, \ and\ \bibinfo {author}
  {\bibfnamefont {G.~R.}\ \bibnamefont {Fleming}},\ }\href@noop {} {\bibfield
  {journal} {\bibinfo  {journal} {Nature}\ }\textbf {\bibinfo {volume} {446}},\
  \bibinfo {pages} {782 } (\bibinfo {year} {2007})}\BibitemShut {NoStop}%
\bibitem [{\citenamefont {Lee}\ \emph {et~al.}(2007)\citenamefont {Lee},
  \citenamefont {Cheng},\ and\ \citenamefont {Fleming}}]{Lee2007}%
  \BibitemOpen
  \bibfield  {author} {\bibinfo {author} {\bibfnamefont {H.}~\bibnamefont
  {Lee}}, \bibinfo {author} {\bibfnamefont {Y.-C.}\ \bibnamefont {Cheng}}, \
  and\ \bibinfo {author} {\bibfnamefont {G.~R.}\ \bibnamefont {Fleming}},\
  }\href@noop {} {\bibfield  {journal} {\bibinfo  {journal} {Science}\ }\textbf
  {\bibinfo {volume} {316}},\ \bibinfo {pages} {1462 } (\bibinfo {year}
  {2007})}\BibitemShut {NoStop}%
\bibitem [{\citenamefont {Mohseni}\ \emph {et~al.}(2008)\citenamefont
  {Mohseni}, \citenamefont {Rebentrost}, \citenamefont {Lloyd},\ and\
  \citenamefont {Aspuru-Guzik}}]{Mohseni2008}%
  \BibitemOpen
  \bibfield  {author} {\bibinfo {author} {\bibfnamefont {M.}~\bibnamefont
  {Mohseni}}, \bibinfo {author} {\bibfnamefont {P.}~\bibnamefont {Rebentrost}},
  \bibinfo {author} {\bibfnamefont {S.}~\bibnamefont {Lloyd}}, \ and\ \bibinfo
  {author} {\bibfnamefont {A.}~\bibnamefont {Aspuru-Guzik}},\ }\href@noop {}
  {\bibfield  {journal} {\bibinfo  {journal} {J. Chem. Phys.}\ }\textbf
  {\bibinfo {volume} {129}},\ \bibinfo {pages} {174106} (\bibinfo {year}
  {2008})}\BibitemShut {NoStop}%
\bibitem [{\citenamefont {Kurzy\'nski}(2008)}]{Kurzynski2008}%
  \BibitemOpen
  \bibfield  {author} {\bibinfo {author} {\bibfnamefont {P.}~\bibnamefont
  {Kurzy\'nski}},\ }\href@noop {} {\bibfield  {journal} {\bibinfo  {journal}
  {Phys. Lett. A}\ }\textbf {\bibinfo {volume} {372}},\ \bibinfo {pages} {6125
  } (\bibinfo {year} {2008})}\BibitemShut {NoStop}%
\bibitem [{\citenamefont {Childs}(2009{\natexlab{b}})}]{Childs2009}%
  \BibitemOpen
  \bibfield  {author} {\bibinfo {author} {\bibfnamefont {A.~M.}\ \bibnamefont
  {Childs}},\ }\href@noop {} {\bibfield  {journal} {\bibinfo  {journal}
  {Commun. Math. Phys.}\ }\textbf {\bibinfo {volume} {294}},\ \bibinfo {pages}
  {581 } (\bibinfo {year} {2009}{\natexlab{b}})}\BibitemShut {NoStop}%
\bibitem [{\citenamefont {Kitagawa}\ \emph {et~al.}(2010)\citenamefont
  {Kitagawa}, \citenamefont {Rudner}, \citenamefont {Berg},\ and\ \citenamefont
  {Demler}}]{Kitagawa2010}%
  \BibitemOpen
  \bibfield  {author} {\bibinfo {author} {\bibfnamefont {T.}~\bibnamefont
  {Kitagawa}}, \bibinfo {author} {\bibfnamefont {M.~S.}\ \bibnamefont
  {Rudner}}, \bibinfo {author} {\bibfnamefont {E.}~\bibnamefont {Berg}}, \ and\
  \bibinfo {author} {\bibfnamefont {E.}~\bibnamefont {Demler}},\ }\href@noop {}
  {\bibfield  {journal} {\bibinfo  {journal} {Phys. Rev. A}\ }\textbf {\bibinfo
  {volume} {82}},\ \bibinfo {pages} {033429} (\bibinfo {year}
  {2010})}\BibitemShut {NoStop}%
\bibitem [{\citenamefont {Schreiber}\ \emph {et~al.}(2010)\citenamefont
  {Schreiber}, \citenamefont {Cassemiro}, \citenamefont {Poto\v{c}ek},
  \citenamefont {G\'abris}, \citenamefont {Mosley}, \citenamefont {Andersson},
  \citenamefont {Jex},\ and\ \citenamefont {Silberhorn}}]{Schreiber2010}%
  \BibitemOpen
  \bibfield  {author} {\bibinfo {author} {\bibfnamefont {A.}~\bibnamefont
  {Schreiber}}, \bibinfo {author} {\bibfnamefont {K.~N.}\ \bibnamefont
  {Cassemiro}}, \bibinfo {author} {\bibfnamefont {V.}~\bibnamefont
  {Poto\v{c}ek}}, \bibinfo {author} {\bibfnamefont {A.}~\bibnamefont
  {G\'abris}}, \bibinfo {author} {\bibfnamefont {P.~J.}\ \bibnamefont
  {Mosley}}, \bibinfo {author} {\bibfnamefont {E.}~\bibnamefont {Andersson}},
  \bibinfo {author} {\bibfnamefont {I.}~\bibnamefont {Jex}}, \ and\ \bibinfo
  {author} {\bibfnamefont {C.}~\bibnamefont {Silberhorn}},\ }\href {\doibase
  10.1103/physrevlett.104.050502} {\bibfield  {journal} {\bibinfo  {journal}
  {Phys. Rev. Lett.}\ }\textbf {\bibinfo {volume} {104}},\ \bibinfo {pages}
  {050502} (\bibinfo {year} {2010})}\BibitemShut {NoStop}%
\bibitem [{\citenamefont {Schreiber}\ \emph {et~al.}(2012)\citenamefont
  {Schreiber}, \citenamefont {G\'abris}, \citenamefont {Rohde}, \citenamefont
  {Laiho}, \citenamefont {\v{S}tefa\v{n}ak}, \citenamefont {Poto\v{c}ek},
  \citenamefont {Hamilton}, \citenamefont {Jex},\ and\ \citenamefont
  {Silberhorn}}]{Schreiber2012}%
  \BibitemOpen
  \bibfield  {author} {\bibinfo {author} {\bibfnamefont {A.}~\bibnamefont
  {Schreiber}}, \bibinfo {author} {\bibfnamefont {A.}~\bibnamefont {G\'abris}},
  \bibinfo {author} {\bibfnamefont {P.~P.}\ \bibnamefont {Rohde}}, \bibinfo
  {author} {\bibfnamefont {K.}~\bibnamefont {Laiho}}, \bibinfo {author}
  {\bibfnamefont {M.}~\bibnamefont {\v{S}tefa\v{n}ak}}, \bibinfo {author}
  {\bibfnamefont {V.}~\bibnamefont {Poto\v{c}ek}}, \bibinfo {author}
  {\bibfnamefont {C.}~\bibnamefont {Hamilton}}, \bibinfo {author}
  {\bibfnamefont {I.}~\bibnamefont {Jex}}, \ and\ \bibinfo {author}
  {\bibfnamefont {C.}~\bibnamefont {Silberhorn}},\ }\href {\doibase
  10.1126/science.1218448} {\bibfield  {journal} {\bibinfo  {journal}
  {Science}\ }\textbf {\bibinfo {volume} {336}},\ \bibinfo {pages} {55 }
  (\bibinfo {year} {2012})}\BibitemShut {NoStop}%
\bibitem [{\citenamefont {Chandrashekar}\ \emph {et~al.}(2010)\citenamefont
  {Chandrashekar}, \citenamefont {Banerjee},\ and\ \citenamefont
  {Srikanth}}]{Chandrashekar2010}%
  \BibitemOpen
  \bibfield  {author} {\bibinfo {author} {\bibfnamefont {C.~M.}\ \bibnamefont
  {Chandrashekar}}, \bibinfo {author} {\bibfnamefont {S.}~\bibnamefont
  {Banerjee}}, \ and\ \bibinfo {author} {\bibfnamefont {R.}~\bibnamefont
  {Srikanth}},\ }\href@noop {} {\bibfield  {journal} {\bibinfo  {journal}
  {Phys. Rev. A}\ }\textbf {\bibinfo {volume} {81}},\ \bibinfo {pages} {062340}
  (\bibinfo {year} {2010})}\BibitemShut {NoStop}%
\bibitem [{\citenamefont {Berry}\ and\ \citenamefont
  {Childs}(2012)}]{Berry2012}%
  \BibitemOpen
  \bibfield  {author} {\bibinfo {author} {\bibfnamefont {D.~W.}\ \bibnamefont
  {Berry}}\ and\ \bibinfo {author} {\bibfnamefont {A.~M.}\ \bibnamefont
  {Childs}},\ }\href@noop {} {\bibfield  {journal} {\bibinfo  {journal} {Q.
  Info. Comp.}\ }\textbf {\bibinfo {volume} {12}},\ \bibinfo {pages} {29}
  (\bibinfo {year} {2012})}\BibitemShut {NoStop}%
\bibitem [{\citenamefont {Kitagawa}(2012)}]{Kitagawa2012}%
  \BibitemOpen
  \bibfield  {author} {\bibinfo {author} {\bibfnamefont {T.}~\bibnamefont
  {Kitagawa}},\ }\href@noop {} {\bibfield  {journal} {\bibinfo  {journal}
  {Quantum Inf. Process.}\ }\textbf {\bibinfo {volume} {11}},\ \bibinfo {pages}
  {1107 } (\bibinfo {year} {2012})}\BibitemShut {NoStop}%
\bibitem [{\citenamefont {Kitagawa}\ \emph {et~al.}(2012)\citenamefont
  {Kitagawa}, \citenamefont {Broome}, \citenamefont {Fedrizzi}, \citenamefont
  {Rudner}, \citenamefont {Berg}, \citenamefont {Kassal}, \citenamefont
  {Aspuru-Guzik}, \citenamefont {Demler},\ and\ \citenamefont
  {White}}]{Kitagawa2012a}%
  \BibitemOpen
  \bibfield  {author} {\bibinfo {author} {\bibfnamefont {T.}~\bibnamefont
  {Kitagawa}}, \bibinfo {author} {\bibfnamefont {M.~A.}\ \bibnamefont
  {Broome}}, \bibinfo {author} {\bibfnamefont {A.}~\bibnamefont {Fedrizzi}},
  \bibinfo {author} {\bibfnamefont {M.~S.}\ \bibnamefont {Rudner}}, \bibinfo
  {author} {\bibfnamefont {E.}~\bibnamefont {Berg}}, \bibinfo {author}
  {\bibfnamefont {I.}~\bibnamefont {Kassal}}, \bibinfo {author} {\bibfnamefont
  {A.}~\bibnamefont {Aspuru-Guzik}}, \bibinfo {author} {\bibfnamefont
  {E.}~\bibnamefont {Demler}}, \ and\ \bibinfo {author} {\bibfnamefont {A.~G.}\
  \bibnamefont {White}},\ }\href@noop {} {\bibfield  {journal} {\bibinfo
  {journal} {Nat. Commun.}\ }\textbf {\bibinfo {volume} {3}},\ \bibinfo {pages}
  {882} (\bibinfo {year} {2012})}\BibitemShut {NoStop}%
\bibitem [{\citenamefont {Asb\'oth}(2012)}]{Asboth2012}%
  \BibitemOpen
  \bibfield  {author} {\bibinfo {author} {\bibfnamefont {J.~K.}\ \bibnamefont
  {Asb\'oth}},\ }\href@noop {} {\bibfield  {journal} {\bibinfo  {journal}
  {Phys. Rev. B}\ }\textbf {\bibinfo {volume} {86}},\ \bibinfo {pages} {195414}
  (\bibinfo {year} {2012})}\BibitemShut {NoStop}%
\bibitem [{\citenamefont {Moulieras}\ \emph {et~al.}(2013)\citenamefont
  {Moulieras}, \citenamefont {Lewenstein},\ and\ \citenamefont
  {Puentes}}]{Moulieras2013}%
  \BibitemOpen
  \bibfield  {author} {\bibinfo {author} {\bibfnamefont {S.}~\bibnamefont
  {Moulieras}}, \bibinfo {author} {\bibfnamefont {M.}~\bibnamefont
  {Lewenstein}}, \ and\ \bibinfo {author} {\bibfnamefont {G.}~\bibnamefont
  {Puentes}},\ }\href@noop {} {\bibfield  {journal} {\bibinfo  {journal} {J.
  Phys. B: At. Mol. Opt. Phys.}\ }\textbf {\bibinfo {volume} {46}},\ \bibinfo
  {pages} {104005} (\bibinfo {year} {2013})}\BibitemShut {NoStop}%
\bibitem [{\citenamefont {Obuse}\ \emph {et~al.}(2015)\citenamefont {Obuse},
  \citenamefont {Asb\'oth}, \citenamefont {Nishimura},\ and\ \citenamefont
  {Kawakami}}]{Obuse2015}%
  \BibitemOpen
  \bibfield  {author} {\bibinfo {author} {\bibfnamefont {H.}~\bibnamefont
  {Obuse}}, \bibinfo {author} {\bibfnamefont {J.~K.}\ \bibnamefont {Asb\'oth}},
  \bibinfo {author} {\bibfnamefont {Y.}~\bibnamefont {Nishimura}}, \ and\
  \bibinfo {author} {\bibfnamefont {N.}~\bibnamefont {Kawakami}},\ }\href@noop
  {} {\bibfield  {journal} {\bibinfo  {journal} {Phys. Rev. B}\ }\textbf
  {\bibinfo {volume} {92}},\ \bibinfo {pages} {045424} (\bibinfo {year}
  {2015})}\BibitemShut {NoStop}%
\bibitem [{\citenamefont {Edge}\ and\ \citenamefont
  {Asb\'oth}(2015)}]{Edge2015}%
  \BibitemOpen
  \bibfield  {author} {\bibinfo {author} {\bibfnamefont {J.~M.}\ \bibnamefont
  {Edge}}\ and\ \bibinfo {author} {\bibfnamefont {J.~K.}\ \bibnamefont
  {Asb\'oth}},\ }\href@noop {} {\bibfield  {journal} {\bibinfo  {journal}
  {Phys. Rev. B}\ }\textbf {\bibinfo {volume} {91}},\ \bibinfo {pages} {104202}
  (\bibinfo {year} {2015})}\BibitemShut {NoStop}%
\bibitem [{\citenamefont {Cardano}\ \emph {et~al.}(2015)\citenamefont
  {Cardano}, \citenamefont {Massa}, \citenamefont {Qassim}, \citenamefont
  {Karimi}, \citenamefont {Slussarenko}, \citenamefont {Paparo}, \citenamefont
  {de~Lisio}, \citenamefont {Sciarrino}, \citenamefont {Santamato},
  \citenamefont {Boyd},\ and\ \citenamefont {Marrucci}}]{Cardano2015}%
  \BibitemOpen
  \bibfield  {author} {\bibinfo {author} {\bibfnamefont {F.}~\bibnamefont
  {Cardano}}, \bibinfo {author} {\bibfnamefont {F.}~\bibnamefont {Massa}},
  \bibinfo {author} {\bibfnamefont {H.}~\bibnamefont {Qassim}}, \bibinfo
  {author} {\bibfnamefont {E.}~\bibnamefont {Karimi}}, \bibinfo {author}
  {\bibfnamefont {S.}~\bibnamefont {Slussarenko}}, \bibinfo {author}
  {\bibfnamefont {D.}~\bibnamefont {Paparo}}, \bibinfo {author} {\bibfnamefont
  {C.}~\bibnamefont {de~Lisio}}, \bibinfo {author} {\bibfnamefont
  {F.}~\bibnamefont {Sciarrino}}, \bibinfo {author} {\bibfnamefont
  {E.}~\bibnamefont {Santamato}}, \bibinfo {author} {\bibfnamefont {R.~W.}\
  \bibnamefont {Boyd}}, \ and\ \bibinfo {author} {\bibfnamefont
  {L.}~\bibnamefont {Marrucci}},\ }\href@noop {} {\bibfield  {journal}
  {\bibinfo  {journal} {Sci. Adv.}\ }\textbf {\bibinfo {volume} {1}},\ \bibinfo
  {pages} {1500087} (\bibinfo {year} {2015})}\BibitemShut {NoStop}%
\bibitem [{\citenamefont {Zhang}\ \emph {et~al.}(2010)\citenamefont {Zhang},
  \citenamefont {Liu}, \citenamefont {Liu}, \citenamefont {Li}, \citenamefont
  {Li},\ and\ \citenamefont {Guo}}]{Zhang2010}%
  \BibitemOpen
  \bibfield  {author} {\bibinfo {author} {\bibfnamefont {P.}~\bibnamefont
  {Zhang}}, \bibinfo {author} {\bibfnamefont {B.-H.}\ \bibnamefont {Liu}},
  \bibinfo {author} {\bibfnamefont {R.-F.}\ \bibnamefont {Liu}}, \bibinfo
  {author} {\bibfnamefont {H.-R.}\ \bibnamefont {Li}}, \bibinfo {author}
  {\bibfnamefont {F.-L.}\ \bibnamefont {Li}}, \ and\ \bibinfo {author}
  {\bibfnamefont {G.-C.}\ \bibnamefont {Guo}},\ }\href {\doibase
  10.1103/physreva.81.052322} {\bibfield  {journal} {\bibinfo  {journal} {Phys.
  Rev. A}\ }\textbf {\bibinfo {volume} {81}},\ \bibinfo {pages} {052322}
  (\bibinfo {year} {2010})}\BibitemShut {NoStop}%
\bibitem [{\citenamefont {Goyal}\ \emph {et~al.}(2013)\citenamefont {Goyal},
  \citenamefont {Roux}, \citenamefont {Forbes},\ and\ \citenamefont
  {Konrad}}]{Goyal2013}%
  \BibitemOpen
  \bibfield  {author} {\bibinfo {author} {\bibfnamefont {S.~K.}\ \bibnamefont
  {Goyal}}, \bibinfo {author} {\bibfnamefont {F.~S.}\ \bibnamefont {Roux}},
  \bibinfo {author} {\bibfnamefont {A.}~\bibnamefont {Forbes}}, \ and\ \bibinfo
  {author} {\bibfnamefont {T.}~\bibnamefont {Konrad}},\ }\href@noop {}
  {\bibfield  {journal} {\bibinfo  {journal} {Phys. Rev. Lett.}\ }\textbf
  {\bibinfo {volume} {110}},\ \bibinfo {pages} {263602} (\bibinfo {year}
  {2013})}\BibitemShut {NoStop}%
\bibitem [{\citenamefont {Goyal}\ \emph {et~al.}(2015)\citenamefont {Goyal},
  \citenamefont {Konrad},\ and\ \citenamefont {Di\'osi}}]{Goyal2015}%
  \BibitemOpen
  \bibfield  {author} {\bibinfo {author} {\bibfnamefont {S.~K.}\ \bibnamefont
  {Goyal}}, \bibinfo {author} {\bibfnamefont {T.}~\bibnamefont {Konrad}}, \
  and\ \bibinfo {author} {\bibfnamefont {L.}~\bibnamefont {Di\'osi}},\
  }\href@noop {} {\bibfield  {journal} {\bibinfo  {journal} {Phys. Lett. A}\
  }\textbf {\bibinfo {volume} {379}},\ \bibinfo {pages} {100 } (\bibinfo {year}
  {2015})}\BibitemShut {NoStop}%
\bibitem [{\citenamefont {Genske}\ \emph {et~al.}(2013)\citenamefont {Genske},
  \citenamefont {Alt}, \citenamefont {Steffen}, \citenamefont {Werner},
  \citenamefont {Werner}, \citenamefont {Meschede},\ and\ \citenamefont
  {Alberti}}]{Genske2013}%
  \BibitemOpen
  \bibfield  {author} {\bibinfo {author} {\bibfnamefont {M.}~\bibnamefont
  {Genske}}, \bibinfo {author} {\bibfnamefont {W.}~\bibnamefont {Alt}},
  \bibinfo {author} {\bibfnamefont {A.}~\bibnamefont {Steffen}}, \bibinfo
  {author} {\bibfnamefont {A.~H.}\ \bibnamefont {Werner}}, \bibinfo {author}
  {\bibfnamefont {R.~F.}\ \bibnamefont {Werner}}, \bibinfo {author}
  {\bibfnamefont {D.}~\bibnamefont {Meschede}}, \ and\ \bibinfo {author}
  {\bibfnamefont {A.}~\bibnamefont {Alberti}},\ }\href@noop {} {\bibfield
  {journal} {\bibinfo  {journal} {Phys. Rev. Lett.}\ }\textbf {\bibinfo
  {volume} {110}},\ \bibinfo {pages} {190601} (\bibinfo {year}
  {2013})}\BibitemShut {NoStop}%
\bibitem [{\citenamefont {Cedzich}\ \emph {et~al.}(2013)\citenamefont
  {Cedzich}, \citenamefont {Ryb\'ar}, \citenamefont {Werner}, \citenamefont
  {Alberti}, \citenamefont {Genske},\ and\ \citenamefont
  {Werner}}]{Cedzich2013}%
  \BibitemOpen
  \bibfield  {author} {\bibinfo {author} {\bibfnamefont {C.}~\bibnamefont
  {Cedzich}}, \bibinfo {author} {\bibfnamefont {T.}~\bibnamefont {Ryb\'ar}},
  \bibinfo {author} {\bibfnamefont {A.~H.}\ \bibnamefont {Werner}}, \bibinfo
  {author} {\bibfnamefont {A.}~\bibnamefont {Alberti}}, \bibinfo {author}
  {\bibfnamefont {M.}~\bibnamefont {Genske}}, \ and\ \bibinfo {author}
  {\bibfnamefont {R.~F.}\ \bibnamefont {Werner}},\ }\href@noop {} {\bibfield
  {journal} {\bibinfo  {journal} {Phys. Rev. Lett.}\ }\textbf {\bibinfo
  {volume} {111}},\ \bibinfo {pages} {160601} (\bibinfo {year}
  {2013})}\BibitemShut {NoStop}%
\bibitem [{\citenamefont {Marrucci}\ \emph {et~al.}(2006)\citenamefont
  {Marrucci}, \citenamefont {Manzo},\ and\ \citenamefont
  {Paparo}}]{Marrucci2006}%
  \BibitemOpen
  \bibfield  {author} {\bibinfo {author} {\bibfnamefont {L.}~\bibnamefont
  {Marrucci}}, \bibinfo {author} {\bibfnamefont {C.}~\bibnamefont {Manzo}}, \
  and\ \bibinfo {author} {\bibfnamefont {D.}~\bibnamefont {Paparo}},\
  }\href@noop {} {\bibfield  {journal} {\bibinfo  {journal} {Phys. Rev. Lett.}\
  }\textbf {\bibinfo {volume} {96}},\ \bibinfo {pages} {163905} (\bibinfo
  {year} {2006})}\BibitemShut {NoStop}%
\bibitem [{\citenamefont {Simon}\ and\ \citenamefont
  {Mukunda}(1989)}]{Simon1989}%
  \BibitemOpen
  \bibfield  {author} {\bibinfo {author} {\bibfnamefont {R.}~\bibnamefont
  {Simon}}\ and\ \bibinfo {author} {\bibfnamefont {N.}~\bibnamefont
  {Mukunda}},\ }\href {\doibase 10.1016/0375-9601(89)90748-2} {\bibfield
  {journal} {\bibinfo  {journal} {Phys. Lett. A}\ }\textbf {\bibinfo {volume}
  {138}},\ \bibinfo {pages} {474 } (\bibinfo {year} {1989})}\BibitemShut
  {NoStop}%
\bibitem [{\citenamefont {Nielsen}\ and\ \citenamefont
  {Chuang}(2010)}]{Nielsen2010}%
  \BibitemOpen
  \bibfield  {author} {\bibinfo {author} {\bibfnamefont {M.~A.}\ \bibnamefont
  {Nielsen}}\ and\ \bibinfo {author} {\bibfnamefont {I.~L.}\ \bibnamefont
  {Chuang}},\ }\href@noop {} {\emph {\bibinfo {title} {Quantum Computation and
  Quantum Information}}}\ (\bibinfo  {publisher} {Cambridge University Press,
  Cambridge},\ \bibinfo {year} {2010})\BibitemShut {NoStop}%
\bibitem [{\citenamefont {Kendon}\ and\ \citenamefont
  {Tregenna}(2003)}]{Kendon2003}%
  \BibitemOpen
  \bibfield  {author} {\bibinfo {author} {\bibfnamefont {V.}~\bibnamefont
  {Kendon}}\ and\ \bibinfo {author} {\bibfnamefont {B.}~\bibnamefont
  {Tregenna}},\ }\href@noop {} {\bibfield  {journal} {\bibinfo  {journal}
  {Phys. Rev. A}\ }\textbf {\bibinfo {volume} {67}},\ \bibinfo {pages} {042315}
  (\bibinfo {year} {2003})}\BibitemShut {NoStop}%
\bibitem [{\citenamefont {Kendon}(2007)}]{Kendon2007}%
  \BibitemOpen
  \bibfield  {author} {\bibinfo {author} {\bibfnamefont {V.}~\bibnamefont
  {Kendon}},\ }\href@noop {} {\bibfield  {journal} {\bibinfo  {journal} {Math.
  Struct. Comput. Sci.}\ }\textbf {\bibinfo {volume} {17}},\ \bibinfo {pages}
  {1169 } (\bibinfo {year} {2007})}\BibitemShut {NoStop}%
\bibitem [{\citenamefont {Broome}\ \emph {et~al.}(2010)\citenamefont {Broome},
  \citenamefont {Fedrizzi}, \citenamefont {Lanyon}, \citenamefont {Kassal},
  \citenamefont {Aspuru-Guzik},\ and\ \citenamefont {White}}]{Broome2010}%
  \BibitemOpen
  \bibfield  {author} {\bibinfo {author} {\bibfnamefont {M.~A.}\ \bibnamefont
  {Broome}}, \bibinfo {author} {\bibfnamefont {A.}~\bibnamefont {Fedrizzi}},
  \bibinfo {author} {\bibfnamefont {B.~P.}\ \bibnamefont {Lanyon}}, \bibinfo
  {author} {\bibfnamefont {I.}~\bibnamefont {Kassal}}, \bibinfo {author}
  {\bibfnamefont {A.}~\bibnamefont {Aspuru-Guzik}}, \ and\ \bibinfo {author}
  {\bibfnamefont {A.~G.}\ \bibnamefont {White}},\ }\href@noop {} {\bibfield
  {journal} {\bibinfo  {journal} {Phys. Rev. Lett.}\ }\textbf {\bibinfo
  {volume} {104}},\ \bibinfo {pages} {153602} (\bibinfo {year}
  {2010})}\BibitemShut {NoStop}%
\bibitem [{\citenamefont {Dalvit}\ \emph {et~al.}(2000)\citenamefont {Dalvit},
  \citenamefont {Dziarmaga},\ and\ \citenamefont {Zurek}}]{Dalvit2000}%
  \BibitemOpen
  \bibfield  {author} {\bibinfo {author} {\bibfnamefont {D.~A.~R.}\
  \bibnamefont {Dalvit}}, \bibinfo {author} {\bibfnamefont {J.}~\bibnamefont
  {Dziarmaga}}, \ and\ \bibinfo {author} {\bibfnamefont {W.~H.}\ \bibnamefont
  {Zurek}},\ }\href@noop {} {\bibfield  {journal} {\bibinfo  {journal} {Phys.
  Rev. A}\ }\textbf {\bibinfo {volume} {62}},\ \bibinfo {pages} {013607}
  (\bibinfo {year} {2000})}\BibitemShut {NoStop}%
\bibitem [{\citenamefont {Padgett}\ and\ \citenamefont
  {Lesso}(1999)}]{Padgett1999}%
  \BibitemOpen
  \bibfield  {author} {\bibinfo {author} {\bibfnamefont {M.~J.}\ \bibnamefont
  {Padgett}}\ and\ \bibinfo {author} {\bibfnamefont {J.~P.}\ \bibnamefont
  {Lesso}},\ }\href@noop {} {\bibfield  {journal} {\bibinfo  {journal} {J. Mod.
  Opt.}\ }\textbf {\bibinfo {volume} {46}},\ \bibinfo {pages} {175 } (\bibinfo
  {year} {1999})}\BibitemShut {NoStop}%
\bibitem [{\citenamefont {Hall}\ \emph {et~al.}(2011)\citenamefont {Hall},
  \citenamefont {Altepeter},\ and\ \citenamefont {Kumar}}]{Hall2011}%
  \BibitemOpen
  \bibfield  {author} {\bibinfo {author} {\bibfnamefont {M.~A.}\ \bibnamefont
  {Hall}}, \bibinfo {author} {\bibfnamefont {J.~B.}\ \bibnamefont {Altepeter}},
  \ and\ \bibinfo {author} {\bibfnamefont {P.}~\bibnamefont {Kumar}},\
  }\href@noop {} {\bibfield  {journal} {\bibinfo  {journal} {New J. Phys.}\
  }\textbf {\bibinfo {volume} {13}},\ \bibinfo {pages} {105004} (\bibinfo
  {year} {2011})}\BibitemShut {NoStop}%
\bibitem [{\citenamefont {Nalbach}\ \emph {et~al.}(2011)\citenamefont
  {Nalbach}, \citenamefont {Braun},\ and\ \citenamefont
  {Thorwart}}]{Nalbach2011}%
  \BibitemOpen
  \bibfield  {author} {\bibinfo {author} {\bibfnamefont {P.}~\bibnamefont
  {Nalbach}}, \bibinfo {author} {\bibfnamefont {D.}~\bibnamefont {Braun}}, \
  and\ \bibinfo {author} {\bibfnamefont {M.}~\bibnamefont {Thorwart}},\
  }\href@noop {} {\bibfield  {journal} {\bibinfo  {journal} {Phys. Rev. E}\
  }\textbf {\bibinfo {volume} {84}},\ \bibinfo {pages} {041926} (\bibinfo
  {year} {2011})}\BibitemShut {NoStop}%
\end{thebibliography}
%

\end{document}